%% file: ms.tex
\documentclass[12pt,preprint]{aastex}

\newcommand{\brsq}{\langle B_R^2/B_\phi^2\rangle}
\newcommand{\Qz}{\langle Q_z \rangle}
\newcommand{\Qy}{\langle Q_y \rangle}
\newcommand{\Qp}{\langle Q_\phi \rangle}

\begin{document}
\title{Assessing Quantitative Results in Accretion Simulations:  From
Local to Global}

\author{John F. Hawley, Xiaoyue Guan}
\affil{Department of Astronomy \\ University of Virginia \\ P.O. Box
400325 \\
Charlottesville, VA 22904-4325}
\email{jh8h@virginia.edu; xg3z@virginia.edu}

\and
\author{Julian H. Krolik}
\affil{Department of Physics and Astronomy\\
Johns Hopkins University\\
Baltimore, MD 21218}
\email{jhk@pha.jhu.edu}

\begin{abstract}

\input abstract

\end{abstract}

\keywords{Black holes - magnetohydrodynamics (MHD) - stars:accretion -
methods:  numerical}

\input introduction

\input numerics

\input shearing

\input globals

\input interpretation

\input conclusions

\section*{Acknowledgements}

This work was partially supported by NASA grant NNX09AD14G and NSF
grant AST-0908869 (JFH) and NSF grant AST-0908336 (JHK).  Some of the
simulations described here were carried out the Ranger system at TACC and
on the Kraken system at NICS, both supported by the NSF.  We thank Jake
Simon, Shane Davis and Jim Stone for sharing data from their shearing
box simulations, and Simon, Stone and Kris Beckwith for comments on
the manuscript.

\bibliographystyle{apj}
\bibliography{hawleyreferences}

\input figures

\end{document}

%% file: abstract.tex
Discretized numerical simulations are a powerful tool for investigation
of nonlinear MHD turbulence in accretion disks.  However, confidence in
their {\it quantitative} predictions requires a demonstration that further
refinement of the spatial gridscale would not result in any significant
change.  This has yet to be accomplished, particularly for global disk
simulations.  In this paper, we combine data from previously published
stratified shearing box simulations and new global disk simulations to
calibrate several quantitative diagnostics by which one can estimate
progress toward numerical convergence.  Using these diagnostics, we find
that the established criterion for an adequate numerical description of
linear growth of the magneto-rotational instability (the number of cells
across a wavelength of the fastest-growing vertical wavenumber mode)
can be extended to a criterion for adequate description of nonlinear MHD
disk turbulence, but the standard required is more stringent.  We also
find that azimuthal resolution, which has not often been extensively
examined in previous studies, can significantly affect the evolution
of the poloidal magnetic field.  We further analyze the comparative
resolution requirements of a small sample of initial magnetic field
geometries; not surprisingly, more complicated initial field geometries
require higher spatial resolution.  Otherwise, they tend to evolve to
qualitatively similar states if evolved for sufficient time.  Applying our
quantitative resolution criteria to a sample of previously published
global simulations, we find that, with perhaps a single exception, they
are significantly under-resolved, and therefore underestimate the magnetic
turbulence and resulting stress levels throughout the accretion flow.

%% file: introduction.tex
\section{Introduction}
\label{sec:introduction}

Numerical simulation is an important tool for understanding the dynamics
and evolution of accretion disks.  In the last decade, the increase
in computational power has made it possible to carry out global disk
simulations of increasing complexity, as the state-of-the-art has risen
from pseudo-Newtonian dynamics \citep{Hawley:2001,Machida:2003}
to full three-dimensional general relativistic physics
\citep{Devilliers:2003,Gammie:2003,Anninos:2005}, most recently
including toy-model thermodynamics \citep{Noble:2009a,Shafee:2008}.
These simulations already have several achievements to their
credit.  They have shown that the magneto-rotational instability
\citep[MRI;][]{Balbus:1991,Balbus:1998} can be effective in providing the
required internal stress. They have shown that disks rapidly evolve to a
near-Keplerian distribution of angular momentum regardless of the initial
angular momentum distribution.  They have demonstrated two mechanisms by
which a large-scale field magnetic field can be attached to a black hole,
either inward transport of truly large-scale field \citep{Beckwith:2009}
or spontaneous, if temporary, inflation of field loops within the
accretion flow \citep{McKinney:2004,Devilliers:2003b,Devilliers:2005}.
If the black hole rotates, these large-scale fields can support Poynting
flux-dominated jets.

Nonetheless, a number of questions remain open regarding the
\textit{quantitative} quality of the predictions these simulations can
make for real black holes in Nature.  Some of these questions have to do
with the adequacy of the physical approximations made, most notably the
still rather crude account of thermodynamics in even the best of them.
Radiation forces, although likely very important in numerous black hole
accretion contexts, likewise remain largely omitted in global simulations.
Others have to do with physical choices that must be made in order
to calculate anything, but about which we have no real knowledge,
particularly the initial strength and structure of the magnetic field.
Confidence in the saturation strength of the field is greatest when it
is statistically time-steady, yet very different from its initial state.
It is therefore desirable to begin with a weak magnetic field and let the
natural dynamics of the situation strengthen it.  Although there have been
discussions in the literature that a certain initial field geometries
are ``most natural" \citep[e.g., multiple loops][]{Shafee:2008},
all such arguments appear to be based on little more than subjective
aesthetic judgments.  We might hope that these choices have little
effect on the outcome, but that must be demonstrated; in the case of
jet-launching, we already know that magnetic field geometry plays a
major role \citep{Beckwith:2008a}.

Other questions have to do with the specifics of the numerical
techniques employed.  Every algorithm has certain strengths and weaknesses,
and their impact depends on the nature of the problem being simulated.
Geometric symmetry conditions may be imposed (e.g., imposing azimuthal
boundary conditions periodic across a wedge rather than around a full
circle in order to economize on computer time) whose consequences cannot
easily be determined {\it a priori}.  Transients due to arbitrary initial
conditions must be eliminated.  And every discrete numerical scheme must
be shown to have fine enough resolution, both spatially and temporally,
to describe adequately the physical processes involved.  This last point
is a special concern for accretion simulations because MHD turbulence is
essential to the story, but true microphysical dissipation
operates on a scale so many orders of magnitude smaller than the disk
scale that no conceivable simulation can hope to resolve it.  It is these
more technical questions that are the province of this paper.

In this paper, we will by no means attempt a full comparison of a broad
range of codes and algorithms.  We will, however, present some comparative
data in the shearing-box context in order to estimate the potential level
of contrast in saturated stress and other properties among codes
utilizing different algorithms run at similar resolutions.

We will also pass over the second question, that of overly-restrictive
symmetry conditions.  We will only note that \cite{Schnittman:2006} showed
that in a full-$2\pi$ simulation there is substantial power in stress
fluctuations on azimuthal scales as long as $\sim 1$~radian.  Given that
finding, wedges of at least a quarter-circle are certainly required.

The third question, how to eliminate transients, poses somewhat different
issues, depending upon whether the context is stratified shearing-boxes
or global disks.  In the former case, it has long been established
\citep{Hawley:1995b} that the principal transients are erased in $\sim
10$~orbits, but the turbulence exhibits significant long-term, chaotic
fluctuation power \citep{Winters:2003}, even over durations as long
as the very longest simulations, $\gtrsim 500$ orbits.  Nonetheless,
despite the long-term variability, it is possible to define time-averages
reasonably well.  We will discuss quantitative definition of transient
removal in global disk simulations more fully in \S~\ref{sec:interp}.
Here it suffices to say that statistical stationarity
in global simulations has a dual meaning.  One sense is the same as
for shearing-boxes: given a surface density and an orbital shear, the
system should achieve a statistical steady-state in the amplitude of the
MHD turbulence.  The other sense is that the local surface density must
have a well-defined mean value over some period of time.  This is more
difficult, and can never be achieved over the entire simulation volume.

The fourth and last question, how to determine whether a given simulation
has adequate resolution to return accurate quantitative results, will
occupy most of our attention.  One expects convergence of the numerical
solution to an exact solution as the grid size goes to zero, i.e.,
$\Delta x \rightarrow 0$.  Since that limit is never attained, convergence
generally refers to the observation that a given quantity approaches some
fixed value as $\Delta x$ is reduced.  Different quantities, of course,
have different convergence rates (in practice, the formal convergence
rate, determined by the order of the scheme, need not be manifest at
a given finite resolution), and individual quantities may themselves
be subject to several different criteria.  The issue of convergence is
further complicated when evolving equations without explicit
resistivity and viscosity.  In that case, dissipation occurs at the
grid scale, and the effective Reynolds number of the system becomes a
function of resolution.

As an illustration, consider the central quantity of accretion physics,
the local stress.  Its saturation depends on the interplay between
several different processes.  First, small magnetic fluctuations are
amplified by the MRI; the spatial resolution must therefore be great
enough that the fastest-growing linear modes---for both poloidal and
toroidal fluctuations---$e$-fold at the correct rate.  Next, nonlinear
couplings between different wave modes must also be described properly.
Adequate resolution for this process depends upon the lengthscale of the
shortest important modes, but it is hard to determine {\it a priori} what
that lengthscale may be.  Finally the turbulence must be dissipated at
scales much less than the stirring scale.  \cite{Lesur:2010}
argue that in feasible simulations, {\it all} resolvable wavelengths are
coupled together by non-linear processes.   They also argue, however,
that in real disks the range of non-local coupling in wavevector space
is much smaller than the dynamic range between the stirring scale and
the physical dissipation scale, permitting the creation of a genuine
inertial range.

Initial conditions may also play a role in determining the necessary
resolution.  Because the fastest-growing MRI wavelength is proportional
to the field strength parallel to the wavevector, the weaker the initial
field, the finer the spatial resolution must be.  In the nonlinear
stage when the magnetic field has a fixed large-scale element, even
modest resolution suffices to describe that portion reasonably well,
although a full description of the turbulence places stronger demands.
On the other hand, when even the largest scales maintained by the physics
are smaller, the minimum necessary resolution becomes correspondingly
finer.  Popular initial conditions for zero net-flux magnetic field
configurations illustrate this effect.  When the initial magnetic field
is a single set of nested poloidal loops, many of these field line loops
become attached to the black hole horizon, forming
a radially large-scale magnetic structure linking the black hole and
the disk.  On the other hand, when the initial field is quadrupolar,
with a pair of nested poloidal loops on either side of the midplane,
these loops are free to shrink in size, demanding a much finer spatial
grid if excessive resistive losses are to be avoided.

For reasons of practicality, simulators tend to choose resolutions
that are as fine as possible while still consistent with their computing
budget.  It is therefore often not feasible to test directly for convergence
by performing a new simulation with more gridzones.  Although some sense
of the rate of convergence can be obtained by computing models at lower
resolution, that procedure would not reveal the failure of the highest
resolution simulation to resolve some important feature---it is entirely
possible that the reason the results do not change appreciably with
changing resolution is that even the best-resolved case is incapable
of sustaining an important physical effect.

Our approach to these problems is to begin with shearing-box calculations,
in which it is generically easier to reach high resolution than in
global disk simulations.  We will search for quantities related to, but
different from, the stress and that scale with resolution in a way that can
be calibrated.  With these in hand, it becomes possible to gain some idea
of how far along the path toward convergence a global simulation has reached.
We will apply these measures both to a number of older simulations and
to some new ones we have carried out especially for this purpose. Full
general relativity is not necessary to achieve these goals because
the saturation of MHD turbulence is an essentially Newtonian process.
Consequently, it is sufficient to carry out simulations using the
simpler and less computationally demanding {\it Zeus} algorithm and a
pseudo-Newtonian potential.

%% file: numerics.tex
\section{Numerical Setup and Initial Conditions}
\label{sec:numerics}

In this study we explore some of the effects of resolution
and initial conditions first by examining the results from local
stratified shearing box simulations, and then by computing a series
of global accretion simulations.  The shearing box results are, in
all but two cases, taken
from the literature.  Our sample features several resolutions, box
sizes, and numerical algorithms.  Some of the numerical aspects of
those simulations are described in \S\ref{sec:shear}; for a detailed
account the reader should consult the papers where these simulations
were originally presented.

Here we describe the numerical setup for the global accretion simulations
carried out specifically for this paper.  As we will be focusing on the
body of the disk itself rather than the interactions with the black hole
or any jets or winds that form, it is sufficient to use non-relativistic
MHD and to work in the pseudo-Newtonian potential to approximate the
gravity of a black hole.  We evolve the equations of Newtonian MHD in
cylindrical coordinates $(R,\phi,z)$,
\begin{equation}\label{mass}
{\partial\rho\over \partial t} + \nabla\cdot (\rho {\bf v}) =  0
\end{equation}
\begin{equation}\label{mom}
\rho {\partial{\bf v} \over \partial t}
+ (\rho {\bf v}\cdot\nabla){\bf v} = -\nabla\left(
P + {\mathcal Q} +{B^2\over 8 \pi} \right)-\rho \nabla \Phi +
\left( {{\bf B}\over 4\pi}\cdot \nabla\right){\bf B}
\end{equation}
\begin{equation}\label{ene}
{\partial\rho\epsilon\over \partial t} + \nabla\cdot (\rho\epsilon
{\bf v}) = -(P+{\mathcal Q}) \nabla \cdot {\bf v}
\end{equation}
\begin{equation} \label{ind}
{\partial{\bf B}\over \partial t} =
\nabla\times\left( {\bf v} \times {\bf B} \right)
\end{equation}
where $\rho$ is the mass density, $\epsilon$ is the specific internal
energy, ${\bf v}$ is the fluid velocity, $P$ is the pressure, $\Phi$
is the gravitational potential, ${\bf B}$ is the magnetic field
vector, and ${\mathcal Q}$ is an explicit artificial viscosity of
the form described by \cite{Stone:1992a}.
To model a black hole gravitational field we use
the pseudo-Newtonian potential of Paczy\'nski \& Wiita (1980) which is
\begin{equation}\label{pwp}
\Phi = - {G M \over r-r_g},
\end{equation}
where $r$ is spherical radius, and $r_g \equiv 2GM/c^2$ is the
``gravitational radius,'' akin to the black hole horizon.   For this
potential, the Keplerian specific angular momentum (i.e., that
corresponding to a circular orbit) is
\begin{equation}\label{pwl}
l_{kep} = (GMr)^{1/2} {r \over r-r_g} ,
\end{equation}
and the angular frequency $\Omega = l/R^2$.  The orbital period at a
radius $r$ is $P_{orb} = 2\pi \Omega^{-1} = 2\pi r^{3/2} (r-r_g)/r$.
The innermost stable circular orbit (ISCO) is located at $r_{ms}=3r_g$.
We use an adiabatic equation of state, $P=\rho\epsilon(\Gamma -1) =
K\rho^\Gamma$, where $P$ is the pressure, $\rho$ is the mass density,
$\epsilon$ is the specific internal energy, $K$ is a constant, and $\Gamma
= 5/3$.  Radiation transport and losses are omitted.  Since there is no
explicit resistivity or physical viscosity, the gas can heat only through
adiabatic compression or by artificial viscosity which acts in shocks,
including weak shocks found in the turbulence.

The code used is a time-explicit Eulerian finite-differencing
{\it Zeus} code for MHD (Stone
\& Norman 1992a; Stone \& Norman 1992b; Hawley \& Stone
1995).\footnote{Here the term {\it Zeus} refers to the algorithm rather
than a specific code implementation.  There are several publicly
available versions of {\it Zeus} as well as a large number of
individually developed versions, such as was employed here.}
We set $GM=c=1$, so that $r_g = 2M$.  Time and distance units are given in
terms of the mass $M$.  

The initial conditions consist of an orbiting gas torus with an angular
momentum distribution parameter $q=1.65$ ($\Omega \propto R^{-q}$),
and $K=0.0034$.  The pressure maximum radius is $R = 35M$ and the inner
torus edge is at $20M$.  The orbital period at the pressure maximum
is $1227M$.  Using the definition of scale height $H^\prime$ in terms
of the density moment, $H^\prime/R = 0.1$ at the pressure maximum.
Explicitly fitting a Gaussian function $\propto \exp\left[-(z/H)^2\right]$
to the vertical pressure distribution gives a value of $H/R \approx 0.16$.
Note that $H = \sqrt{2}H_G$, where $H_G$ is more commonly regarded as the
Gaussian scale height.  In an isothermal thin disk, $H_G = c_s/\Omega$.
In internal code units the initial mass of the torus is 6096 (assuming
a full $2\pi$ azimuthal domain;  the actual computational domain runs
from 0 to $\pi/2$).  Of this total mass, 19\% lies inside of $R=35M$.

We use two initial field configurations and two average field strengths.
The first is the standard dipole loop, in which the vector potential is
written in the form
\begin{equation}
A_\phi = C (\rho - \rho_{cut}) 
\end{equation}
so that the field lines run along contours of constant density.  Here
$C$ is a constant that sets the overall field strength and is set to
zero wherever $\rho < \rho_{cut}$.  This configuration is referred to
as ``one-loop."
We initialize a ``two-loop'' simulation using the vector potential
function of \cite{Shafee:2008}, namely
\begin{equation}
A_\phi = \left[ (\rho - \rho_{cut}) r^{0.75}\right]^2 
\sin\left[\ln (r/S)/T\right]
\end{equation}
where $r$ is the spherical coordinate radius, $\rho_{cut}$ is set
at 20\% of the density maximum (thus confining the initial field to
well within the edge of the initial torus), $S = 1.1 r_{in}$, where
$r_{in}=20 M$ is the initial inner edge of the torus, and $T=0.16$
(see Fig.~\ref{fig:initialfield}).  For either geometry, the initial
field strength is normalized to a $\beta$ value, either 100 or 1000,
defined as the ratio of the total volume-integrated gas pressure to the
volume-integrated magnetic pressure.

Boundary conditions are periodic in $\phi$ and outflow along the $z$
and $R$ boundaries.  One consequence of cylindrical coordinates and
our use of an inner boundary radius located at $R_{in} > r_g$, is that
there is a cutout parallel to the z-axis through which matter as well as
field can pass and leave the grid.  Therefore, in these simulations no
large-scale\ field can become attached to the black hole or fill the axial
region as is seen in the GR simulations.  Because the focus here is on
the evolution of the disk away from the black hole and plunging region,
we accept this limitation for the sake of concentrating computational
power on the torus itself.

The grid used in this study is designed to place as many zones as possible
within the main body of the accretion disk. 
The inner boundary of the radial grid
is set at $R_{in}=4M$. From there it runs outward with constant $\Delta R$ to
$R_1$, beyond which $\Delta R/R$ is set to a constant.  This definition
produces a logarithmically stretched grid throughout most of the domain
while avoiding overly small $\Delta R$ values near the inner boundary.
The $z$ grid concentrates half of the total number of zones symmetrically
around the equator using equal $\Delta z$.  This portion of the grid
extends to $z = \pm z_1$ from the equator.  The remaining $z$ zones are
logarithmically stretched outward to boundaries that are well-removed
from the initial torus.  The $\phi$ grid uses evenly spaced zones that
span one quarter of the full $2\pi$.

We choose a fiducial grid, Grid M (``medium resolution'') and vary the
resolution relative to it.  Grid M contains 256 radial zones with 48
zones inside of $R_1=20M$.  The outer radial boundary is at $R=253M$.
The $z$ grid uses 288 zones with 144 grid zones covering the range
$|z|\leq z_1 = 5M$.  Using the density moment definition of the scale
height, $H^\prime \sim 3.3 M$ at the location of the pressure maximum,
giving $\sim 48$ $z$ zones per $H^\prime$.  Outside of $z = \pm 5 M$
the grid is logarithmically stretched out to $z=\pm 54 M$.  The $\phi$
grid uses 64 equally-spaced zones.  In the inner disk (i.e., $6M \le R
\le 20M$ and $|z| \le 5 M$), the cell aspect ratio $\Delta R/\Delta z =
4.8$ and $\Delta R/(R\Delta\phi) = 2.3$--0.68 (from $R=6M$ to $R=20M$).

The simulations performed on Grid M will be contrasted with the results
obtained using other grid resolutions.  The lower resolution Grid L (``low
resolution'') is half as well resolved in $R$ and $z$.  It consists of
128 radial zones with 24 equally spaced zones between $R=4M$ and $20M$,
with $\Delta R$ increasing $\propto R$ from that radius out to $R=253M$.
At the pressure maximum, $\Delta R = 1.07 M$.  The $z$ grid has 144 zones,
half of which are concentrated inside $z=\pm 5 M$, where the minimum
$\Delta z$ is $0.139 M$.  The $\phi$ grid is the same as in Grid M.
In this case, $\Delta R/\Delta z = 4.8$ as for Grid M, but $\Delta
R/(R\Delta\phi) = 4.5$--1.4.

Grid R (``high-R resolution'') is the same as Grid M for $z$ and $\phi$,
but the number of radial zones is increased to 816.  Inside $R_1 = 10M$,
$\Delta R = 0.04 M$; outside that radius, $\Delta R/R = 0.004$.
The outer boundary of the radial grid is located at $R=142M$.  Its cell
aspect ratios are $\Delta R/\Delta z = 0.58$--1.2 (once again, from
$R=6M$ to $R=20M$) and $\Delta R/(R\Delta\phi) = 0.27$--0.16.

Two other grids are used to study the influence of the $\phi$ resolution.
In Grid PL, the number of $\phi$ zones is reduced to 32.  It uses 256 $R$
zones, but redistributes them to decrease $\Delta R$.  For this grid,
$\Delta R = 0.125 M$ inside of $R=10M$, and $\Delta R/R = 0.0124 M$ outside
that point.  Grid PH is the same as Grid M but with the number of $\phi$
zones increased to 128.

We employ a variety of diagnostics to analyze the simulations.  Certain
quantities, such as total magnetic and kinetic energies, are computed and
recorded every 10 timesteps.  We also save binary data files and compute
integrals of quantities over cylindrical shells at regular time intervals.

%% file: shearing.tex
\section{Shearing Boxes and Resolution Diagnostics}\label{sec:shearing}
\label{sec:shear}

To quantify the resolution effects in global simulations we
begin with the shearing box \citep{Hawley:1995b}, a system where
greater effective resolution can be employed and more extensive
resolution studies are possible.  Since shearing box models
were first introduced, many studies have been carried out to
measure how the stress levels depend on resolution, the initial
field strength and geometry, and other thermodynamic factors
\citep[e.g.,][]{Hawley:1995b,Hawley:1996,Brandenburg:1995,Stone:1996,
Fleming:2000,Miller:2000, Sano:2001,Sano:2002,Sano:2004,Fromang:2007,
Simon:2009a} The first shearing box simulations employed comparatively few
total zones (e.g., $32\times 64\times 32$ zones in $x,y,z$), but recently
the number of zones used in simulations has substantially increased.

Resolution studies using stratified shearing boxes that
include the vertical component of gravity are the most relevant
for comparison to global simulations.  We have examined data
from several recent stratified shearing box resolution studies
\citep{Simon:2011,Shi:2010,Davis:2010,Guan:2011} to see what, if
any, general properties of the MRI turbulence might prove sensitive
to resolution.  Although shearing box simulations have found that
the magnitude of a physical resistivity and viscosity as well as
their ratios can have a significant impact on the turbulence levels
\citep[e.g.][]{Fromang:2007b,Lesur:2007}, at least for smaller magnetic
Reynolds numbers \citep{Oishi:2011}, nearly all global simulations done
to date have used only gridscale dissipation.  Therefore, we consider
here only those shearing box simulations without explicit (physical)
small-scale dissipation.  The resolutions in these simulations range
from 16 to 128 zones per scale height $H$.  The Simon et al. and Davis et
al. models have an isothermal equation of state and make use of the {\it
Athena} code with the HLLD-flux solver \citep{Stone:2008, Stone:2010},
but differ in regard to the initial magnetic field and box size.
The Davis models use a zero-net vertical field initial condition and
a box that is $H:4H:4H$ $(x:y:z)$, while the Simon models have a net
toroidal field overlaid with a poloidal field loop and a box that is
$2H:4H:8H$ $(x:y:z)$; in both cases, $H \equiv \sqrt{2}c_s/\Omega$.
The Shi et~al. models have the same initial field configuration as Simon,
but use a {\it Zeus} code and include heating and radiative transport.
Their box had side lengths $2H^\prime:8H^\prime:16H^\prime$ (x:y:z).
The \cite{Guan:2011} models use a {\it Zeus} code, an isothermal equation
of state, and focus on larger stratified boxes; their fiducial simulation
has a domain of $16H:20H:10H$.  Table \ref{table:shearingbox} presents
some time-averaged data collected from \cite{Simon:2011}, \cite{Shi:2010},
\cite{Davis:2010}, and \cite{Guan:2011} along with two unpublished 8
and 16 zones/$H$ simulations (Simon, private communication).  Note that
the Shi et~al. data use a time-averaged scale height
because the temperature in their simulations varies
as determined by thermal balance between dissipation of the turbulence
and radiative cooling.

It is immediately clear that resolution can strongly affect the quantity
most important to disk evolution, the magnetic stress.  Results from four
stratified shearing box simulations that use 64, 32, 16 and 8 zones per
$H$ for the first 150 orbits are shown in Figure~\ref{fig:shrstress}.
The plotted quantity is the volume-averaged stress parameter $\alpha$,
defined 
\begin{equation} \alpha = {\langle(\rho v_x \delta v_y - B_x
B_y/4\pi )\rangle
         \over \langle \rho \rangle c_s^2}
\end{equation} 
The 64 and 32 zone simulations are the 64Num and 32Num models of
\cite{Simon:2011}; the 16 and 8 zone simulations are lower resolution
versions of those simulations (Simon, private communication).  All four
simulations show initial growth to a peak that occurs at around 10
orbits in time.  After this point, the 8 zone simulation (the lowest
curve) clearly dies out.  Large temporal variations characterize
the other simulations.  The stress in the 16 zone run, for example,
declines slowly for the first 50 orbits, but then rises by a factor of 3
by orbit 60.  The 32 and 64 zone runs maintain a time-averaged $\alpha$
that is consistent with the initial peak value.  The 16 zone run also
varies strongly with time, but with a mean $alpha$ that is less than in
the 32 zone run.  \cite{Simon:2011} note that in unstratified simulations
the time-averaged $\alpha$ changes by a larger amount in going from 16
to 32 zone resolution than in going from 32 to 64 zones per $H$ (their
Figure 1).  In both the Davis et al. and Shi et al. resolution studies,
the smallest number of cells per scale height was $\simeq 20$--30, and
there is little change in $\alpha$ when finer resolutions are employed,
including the best-resolved Davis et al. simulation, in which there are
128 cells per scale height.  In \cite{Guan:2011} the value of $\alpha$
went from $0.013$ to $0.023$ when the resolution of their fiducial model
was doubled,  a change from 13 to 26 zones per $H$
in the vertical direction.

The converged value of $\alpha$ in these particular stratified
isothermal shearing boxes is of order $\simeq 0.02$.  There are,
of course, additional effects beyond resolution that determine
$\alpha$.  As discussed above, resistivity and viscosity can have a
significant impact.  At the same resolution, Simon
et~al. find consistently larger values of the stress compared
to Davis et al., sometimes by close to a factor of two.
The Simon et~al. box is a factor of two larger in both the $x$ and
$z$ dimensions.  The \cite{Guan:2011} simulations also provide some
evidence that $\alpha$ can be larger when larger domains are used.
There is evidence from unstratified shearing boxes that taller boxes also
promote stronger magnetic field (Stone, private communication).  The Shi
et~al. simulations, in which the equation of state directly balances
heating and radiative cooling, suggest that $\alpha$ may be somewhat
larger when more realistic thermodynamics are employed.

Even using the local shearing box, few simulations are carried out with
resolutions as fine as 128 cells per scale height, so in practice one
often asks the question, ``In this particular simulation with only modest
resolution, how close is the measured value of stress to the numerically
converged value?"  Because high-resolution simulations can be very
expensive in computer time, it is useful to define metrics of simulation
quality that can be calibrated to a set of standard simulations and then
applied to the data of a new simulation.  In this way, how close that
simulation comes to convergence may be estimated without the expense
of additional, higher resolution simulations.  In the remainder of this
section, we discuss several such quantities.

\subsection{Convergence metric \#1: $Q_z$}

The first such metric comes from the linear theory of the MRI.
\cite{Noble:2010} used the vertical field characteristic
MRI wavelength to compute a quality parameter $Q_z$ defined by
\begin{equation}
Q_z = \lambda_{MRI} / \Delta z = \frac{2 \pi |v_{az}|}{\Omega \Delta z} ,
\end{equation}
where $v_{az}$ is the $z$ component of the Alfv\'en speed.  The 
characteristic
wavelength $\lambda_{MRI}$ is close to,
but not precisely equal to, the fastest growing MRI wavelength.
Wavelengths $\lambda < \lambda_{MRI}/\sqrt 3 $ are stable, while
all wavelengths $\lambda > \lambda_{MRI}$ are unstable, 
albeit with reduced growth rates $\propto ({\bf k} \cdot {\bf v_{az}})$.  On the
basis of unstratified shearing box simulations, \cite{Sano:2004} suggested
that a $Q_z$ value greater than 6 was required in order to achieve a
linear growth rate close to the analytic prediction.  Considering an
isothermal thin disk with only vertical field in the initial condition,
$\lambda_{MRI}$ can be rewritten in terms of the plasma $\beta$ by noting
that $\beta = 2 \rho H^2 \Omega^2 / B^2$, and
hence $\lambda_{MRI} = 2\pi H \beta^{-1/2}$.  Thus, a value of $Q_z$
of $\sim 10$ requires $1.6 \beta^{1/2}$ zones per $H$ when the 
field is purely vertical;
when the field has any other sort of geometry, $\beta$ in this expression
should be scaled by the fraction of the field energy in the vertical
component, giving a zone total of
\begin{equation}\label{eqn:qz}
N_z \simeq 16 \left(\beta/100\right)^{1/2} 
\left(\langle v_A^2\rangle/\langle v_{Az}^2\rangle\right)^{1/2} \left(Q_z/10\right)
\end{equation}
per scale height $H$.  Because the fraction of the magnetic
energy in vertical field is often only $\sim 0.01$--$0.1$, the number
of zones required for a given $\beta$ increases by $\sim 3$--$10$.

The second column of Table~\ref{table:shearingbox} shows the values
of $Q_z$, averaged over the midplane region ($|z| \leq 0.5H$) for our
stratified shearing box simulation sample.  It is necessary to pick out
the midplane region because $|v_{Az}|$ generically increases sharply
away from the midplane.  Consequently, in these simulations in which the
vertical resolution is uniform (unlike typical global simulations), $Q_z$
generally increases by 1--2 orders of magnitude from $z=0$ to $z \simeq
3H$.  These regions with better effective resolution can be important in
maintaining the turbulence.  By $|z| =2$--$3H$, $\beta \le 1$, and the MRI
is largely suppressed and the large values of $Q_z$ are less relevant.
Comparing the Davis et~al. series with the Simon series, we see that
even with 32 cells per scale height, the Sano et~al.  criterion is met
only marginally.  Although $\alpha$ increases dramatically when $Q_z$
rises past a few, its dependence on resolution (in relative terms)
appears to level out in the range $10 \lesssim Q_z \lesssim 20$.

\subsection{Convergence metric \#2: $Q_y$}

Maintenance of poloidal field and turbulence requires non-axisymmetric
motion.  To estimate how well non-axisymmetric stirring is described
by the simulation, we can define a merit parameter $Q_y$ based on
the toroidal field and the $y$ grid zone size ($Q_\phi$ and $R\Delta\phi$
for global simulations).  The toroidal field MRI is
nonaxisymmetric, and the linear properties of those nonaxisymmetric
modes are somewhat different from those of the vertical field MRI.
Although the nonaxisymmetric MRI modes depend on toroidal field, the
presence of weak poloidal components can greatly increase the
total amplification of nonaxisymmetric modes beyond what is predicted for
a purely toroidal field \citep{Balbus:1992}.  Like the case of vertical
wavevectors, the maximum linear growth rate occurs for wavelengths
comparable to the distance an Alfv\'en wave travels in one orbit, but mode
growth also depends on the radial wavelength, which evolves due to shear.
Further, maximum growth also demands vertical wavenumbers $k_z$ much
greater than $H^{-1}$ \citep{Balbus:1992}.  For shearing box
simulations, the number of $y$ zones required to achieve $Q_y \sim 10$ is
\begin{equation}
N_y \sim 64 \left( H/4\right) \left( \beta /100\right)^{1/2}\left(Q_y/10\right)
\end{equation}
for a $y$ direction spanning $4H$.  For toroidal modes in global
simulations, $Q_\phi = 2\pi H/(\beta^{1/2} R\Delta \phi)$, where $\beta$
includes only the toroidal field component.  To resolve
linear growth of the toroidal MRI in a full $2\pi$ simulation requires
\begin{equation}\label{eqn:qy}
N_\phi \simeq 1000~(0.1~R/H)~(\beta/100)^{1/2}~(Q_\phi/10)
\end{equation}
azimuthal cells.

The simulations described in Table~\ref{table:shearingbox} are nearly
all well-resolved by the $Q_y$ criterion; this is one of the advantages
of shearing boxes.  Only in one case (the Simon 8 cells per $H$ run) is
$Q_y < 10$.  In both the Davis et~al. and Simon et~al. simulations, the
cells are cubical.  Because shear ensures that the azimuthal component
of the magnetic field is much stronger than the vertical component,
cell sizes that are too coarse to yield good vertical resolution can 
nonetheless be quite adequate to describe azimuthal behavior.  However,
in simulations whose grids are elongated in the azimuthal direction,
$Q_y$ values will be smaller.  In the better-resolved Davis and
Simon simulations for example, $Q_y/Q_z \sim 4$ and one might expect that
if $\Delta y/\Delta z \sim 4$, $Q_y$ would be only comparable to $Q_z$.

\subsection{Convergence metric \#3: $\alpha_{mag}$}

Whereas $Q_z$ and $Q_y$ derive from the critical linear wavelength of the
MRI, the other diagnostics we have studied are more closely related to
{\it nonlinear} development of the MHD turbulence.  That is, they reflect
how well a numerical calculation replicates macroscopic magnetic field
properties related to the stress that are independent of discretization
(i.e., convergence).

The first of the nonlinear diagnostics is the ratio of the
Maxwell stress to the magnetic pressure, defined
$\alpha_{mag} = -2 B_R B_\phi / B^2$.  Although turbulent
Reynolds stress also contributes to angular momentum transport at a level
roughly 1/4 of the Maxwell stress, it is difficult to quantify in global
simulations, so we do not include it in our definition of $\alpha_{mag}$.
Even the earliest shearing box simulations \citep{Hawley:1995b} found
that this quantity was remarkably constant from simulation
to simulation.  More recently, \cite{Blackman:2008} examined
a large sample of published unstratified shearing box results
and found that, quite generally, $\alpha \beta \propto \alpha_{mag}$
is roughly constant, where $\alpha$ is the traditional constant of
proportionality between stress and (gas) pressure; the combination
$\alpha\beta$ simply removes the gas pressure from consideration.
Recomputing the results from \cite{Blackman:2008} in terms of our
definition of $\alpha_{mag}$ (i.e., without the Reynolds stress), we find
that their derived values are more or less consistent with $\alpha_{mag}
= 0.3$--$0.4$ as seen in the shearing box simulations reported here,
except for the one with only 8 cells per $H$.

The values of $\alpha_{mag}$ presented in Table~\ref{table:shearingbox}
are derived by taking the ratio of the Maxwell stress integrated over the
central scale height to the similarly integrated magnetic pressure and
then time-averaging; this is how $\alpha_{mag}$ has been determined in
past shearing box simulations.  Other averaging
procedures, such as averaging the local ratio rather than taking the
ratio of the averages, give values that are generally somewhat smaller.
The data show that at a gross level, $\alpha_{mag}$ and $\alpha$ are
correlated: a very low value of $\alpha_{mag}$ corresponds to a very low
value of $\alpha$.  In the lowest resolution simulation discussed here,
with only 8 cells per vertical scale height, $\alpha_{mag} =0.08$.
However, with an even modest improvement in resolution, both $\alpha$
and $\alpha_{mag}$ rise.  

The value of $\alpha_{mag}$ is relatively constant because in
MRI turbulence $B_x$ and $B_y$ are highly correlated.  Both the
background shear, which creates toroidal field out of radial, and the
action of the MRI itself, which stretches out radial field as angular
momentum is transferred between fluid elements, create this correlation.
Time-averaging and volume-averaging over the central scale height, we find
that the Pearson correlation coefficient $C(B_x,B_y)$ is -0.73, -0.70
and -0.67 for the 32, 64, and 128 zones per $H$ Davis runs, and -0.28,
-0.75, -0.73, and -0.71 for the 8 through 64 zones per $H$ Simon runs.
In other words, once one is past a resolution threshold (between 8 and
16 zones per scale height), the correlation rapidly achieves a value
$\simeq -0.7$.  We have computed the correlation averaged over the $(x,y)$
plane as a function of $z$.  For $|z| <2$ it is consistently $\sim -0.7$,
but approaches zero at higher altitudes.  Two scale heights from the
midplane is where the stress dies out \citep{Simon:2011,Guan:2011}.

We can probe a bit deeper into the nature of $\alpha_{mag}$ by
studying its probability distribution over the set of grid cells.
Figure~\ref{fig:alphamag} shows a time-averaged distribution function
for $\alpha_{mag}$ for the region within 2 scale heights of the equator
for each of the four Simon runs.  The low resolution distribution
function peaks around $0$; there is very little net stress remaining
within the decaying turbulence, even while the azimuthal field persists.
As resolution increases, the distribution shifts to higher values, as does
the mean value.  On a zone-to-zone basis, $\alpha_{mag}$ is correlated
with $P_{mag}$ in the sense that where the field is particularly strong,
the ratio of stress to magnetic pressure is also particularly high.
In order of increasing resolution, the mean $\alpha_{mag}$ values are
0.057, 0.275, 0.346 and 0.380.  Thus, improving resolution also leads to
greater correlation between $B_x$ and $B_y$, but the average $\alpha_{\rm
mag}$ saturates at $\sim 0.4$.

Before leaving this topic, we note that the resolution-dependence of
$\alpha_{mag}$ illustrates an important aspect of convergence-testing.
If one compared its behavior in the 8 zone run with simulations having
fewer cells per scale height, one might have concluded that $\alpha_{mag}$
is always $\ll 1$.  In other words, low resolution simulations can be
entirely blind to important effects, and convergence does not even begin
until a resolution threshold is reached where that effect is at least
minimally resolved.

\subsection{Convergence metrics \#4 and \#5: $\langle B_x^2/B_y^2\rangle$ and
$\langle B_z^2/B_x^2\rangle$}

The $\alpha_{mag}$ parameter depends, in part, on the relative magnitude
of the poloidal and toroidal magnetic field components and in part on the
degree of correlation between the radial and toroidal components.  As we
have already seen in our discussion of the $Q_z$ and $Q_y$ diagnostics,
the fidelity with which a given simulation follows the linear growth of
these two field components can be different.  The same may be true of
their nonlinear characteristics.

To explore this question in a way that focuses on the separate
components, independent of their correlation, we examined the time- and
volume-averaged energy ratios $\langle B_x^2/B_y^2\rangle$ and $\langle
B_z^2/B_x^2\rangle$ as functions of resolution.  A relatively clear trend
emerges from the former.  Figure~\ref{fig:shrbxby} shows the time history
of this ratio for the simulations of \cite{Simon:2011}.  It increases
from $< 0.01$ to $0.15$--$0.18$ from the worst to the best resolved models.
Despite the substantial time variability, at every stage the curves in this figure are
well-separated from each other, demonstrating a systematic increase of this
parameter with resolution.  When one looks at the full ensemble of shearing
box simulations, the dependence on resolution becomes even more striking
(Fig.~\ref{fig:bxsqbysq}), with the value apparently leveling off near
0.2 at the highest resolution, when there are at least $\simeq 40$ cells
per $H$.  On the other hand there seems to be
no general trend for the ratio of vertical to radial field energies,
$\langle B_z^2/B_x^2\rangle$; the values range from $\sim 0.4$--$0.6$
in these simulations.

Moreover, as we will discuss in greater detail later, there is
a strong correlation between the $Q$ values and
$\langle B_x^2/B_y^2\rangle$.  The simulations achieving
near-saturation values of $\langle B_x^2/B_y^2\rangle$ (Simon64, Davis64,
Davis128, and ShiDBLE) all have $Q_z \ge 10$ and $Q_y \ge 32$.

\begin{deluxetable}{l|cccccccc}
\tabletypesize{\scriptsize}
\tablewidth{0pc}
\tablecaption{Shearing Box Simulations\label{table:shearingbox}}
\tablehead{
 \colhead{Reference}&
 \colhead{Zones$/H$}&
 \colhead{$Q_z$}&
 \colhead{$Q_y$}&
 \colhead{$B_x^2/B_y^2$}&
 \colhead{$B_z^2/B_x^2$}&
 \colhead{$\beta^{-1}$}&
 \colhead{$\alpha_{mag}$}&
 \colhead{$\alpha$}
}
\startdata
Simon8 (unpub)& 8   & 0.08 & 1.7 & 0.016 & 1.0  & 0.015 & 0.08 & 0.001\\
Simon16 (unpub)& 16  & 2.0  & 13. & 0.075 & 0.53 & 0.057 & 0.30 & 0.016\\
Simon32      & 32  & 5.7  & 27. & 0.13  & 0.53 & 0.072 & 0.37 & 0.025\\
Simon64      & 64  & 11.  & 44. & 0.17  & 0.53 & 0.056 & 0.40 & 0.020\\
Davis32      & 32  & 4.5  & 23. & 0.12  & 0.41 & 0.078 & 0.33 & 0.020 \\
Davis64      & 64  & 10.  & 40. & 0.16  & 0.47 & 0.051 & 0.36 & 0.012\\
Davis128     & 128 & 26.  & 98. & 0.18  & 0.50 & 0.053 & 0.36 & 0.018 \\
ShiSTD       & 27  & 4.8  & 13. & 0.10  & 0.65 & 0.075 & 0.27 & 0.020 \\
ShiZ512      & 53  & 11.  & 13. & 0.12  & 0.56 & 0.130 & 0.30 & 0.029 \\
ShiDBLE      & 50  & 15.  & 32. & 0.15  & 0.63 & 0.098 & 0.22 & 0.029 \\
Guan std16   &12.8 & 2.6  & 15. & 0.07  & 0.58 & 0.035 & 0.28 & 0.013 \\
Guan s16a    &25.6 & 6.8  & 34. & 0.12  & 0.58 & 0.057 & 0.32 & 0.023 \\
\enddata
\end{deluxetable}

\subsection{Summary: Stratified Shearing Box} 

In principle, a resolution study is directly applicable only to a set
of simulations using the same numerical scheme on the same problem.
Only after cross-comparison of parallel resolution studies can different
algorithms be calibrated relative to one another.  Previous comparisons
between the {\it Athena} and {\it Zeus} simulations have indicated that
{\it Athena} has lower turbulence decay rates for two-dimensional shearing
sheet simulations compared to {\it Zeus} \citep{Stone:2010, Stone:2009}.
\cite{Stone:2010} suggest that this is due to the use of third-order,
rather than second-order spatial interpolation in {\it Athena} as well
as the use of the HLLD-flux solver.  In the
studies gathered here, the contrast between {\it Zeus} and {\it Athena}
is less obvious: the Shi et~al. simulations and the \cite{Guan:2011}
{\it Zeus} models have diagnostic values comparable to equivalently
resolved {\it Athena} runs.

Thus, from these simulations, we can conclude that for both {\it Athena}
and {\it Zeus} simulations, stratified shearing boxes begin approaching
convergence when resolution is around 40 zones per $H$.  The quantitative
changes that result in increasing the number grid zones beyond this
point are noticeable, but small, compared to the decrease in zone size,
i.e. convergence is occuring at a rate faster than linear in $\Delta x$.
Adequate resolution requires both $Q_z$ and $Q_y$ to be sufficiently
large, but an especially large value of one can somewhat compensate for
a smaller value of the other.  When $Q_y \gtrsim 20$, $Q_z \gtrsim 10$
suffices; however, when $Q_y$ is smaller, $Q_z \gtrsim 15$ is required.
A ratio of radial to toroidal magnetic energy greater than $\ge 0.15$
and $\alpha_{mag} \simeq 0.3$--0.4 are signatures of well-developed
MRI-driven magnetic turbulence.\footnote{These values are obtained from
shearing boxes without a net vertical field or applied resistivity or
viscosity, as appropriate for the global simulations to be discussed here.
Further work is required to characterize shearing box turbulence in the
presence of net vertical field and non-ideal MHD effects}  The ratio
of the vertical to radial magnetic field energy, on the other hand,
shows no particular trend with respect to resolution.

It is likely that the importance of $Q_y$ stems from the essential role
played by nonaxisymmetric processes in maintaining poloidal field energy
and turbulence.  Purely toroidal fields can support an active MRI-driven
turbulence with relatively small vertical fields, but the vertical field
MRI will necessarily generate toroidal field and nonaxisymmetric motions
are essential to maintaining the poloidal field.

Moving in the direction of {\it reduced} resolution, we note that
even if the characteristic MRI wavelength is unresolved (small $Q$)
longer wavelengths will be unstable, albeit with smaller growth rates.
Thus, lower resolution simulations can still have MRI-induced turbulence
and stress, but at correspondingly reduced levels.  Such a state will
be indicated by smaller relative values of the convergence metrics.
Of course, the results from the 8 zone per $H$ simulation show that
there is a resolution limit beyond which turbulence cannot be sustained.

%% file: globals.tex
\section{Global Simulations}

In this section we describe the results from a set of pseudo-Newtonian
global disk simulations intended to investigate the influences of grid
resolution and initial magnetic field strength and topology.  What they
show about progress toward numerical convergence will be discussed in
the following section.

Our parameter study is centered around a fiducial model, designated
twoloop-1000-mr, a simulation with initial $\beta = 1000$, a 
two-loop initial magnetic field, and grid M, the medium resolution grid.
Using the same initial condition, we have explored the effects of both
increasing and decreasing the grid resolution.  To study the effects
of different initial magnetic field configurations, we have also
used the M grid for a simulation with a single dipolar loop initial
condition and the same initial $\beta$, as well as both single and
double dipolar loops with initial $\beta =100$.  
The various models are listed in Table~\ref{table:list}.

\begin{deluxetable}{l|cccc}
\tabletypesize{\scriptsize}
\tablewidth{0pc}
\tablecaption{Global Simulations\label{table:list}}
\tablehead{
 \colhead{Name}&
 \colhead{Grid}&
 \colhead{Type}&
 \colhead{Initial $\beta$}&
 \colhead{Duration ($M$)}
}
\startdata
{\bf twoloop-1000-mr} & M &  2-Loop & 1000 & 66000 \\
twoloop-100-mr  & M &  2-Loop & 100 & 41000 \\
twoloop-1000-lr & L &  2-Loop & 1000 & 41000 \\
twoloop-1000-hr & R &  2-Loop & 1000 & 34000 \\
twoloop-1000-mlp & PL &  2-Loop & 1000 & 40000 \\
twoloop-1000-mhp & PH &  2-Loop & 1000 & 25000 \\
oneloop-100-lr & L & Dipole 1-Loop & 100 & 40000 \\
oneloop-100-mr   & M & Dipole 1-Loop & 100 & 39000 \\
oneloop-1000-mr  & M & Dipole 1-Loop & 1000 & 38000 \\
\enddata
\end{deluxetable}

\subsection{The Fiducial Run}
\label{fiducial}

The $\beta$ value given for a torus is defined as the ratio of the total
thermal to total magnetic pressure.  Although the mean
initial $\beta = 1000$, the $\beta$ value at any
given point within the torus varies. The minimum value is $\beta =283$
and occurs along the radial field lines above and below the equator in
the inner field loop; where the two loops meet at the pressure maximum,
$\beta = 412$.  For those regions of the torus where there is no magnetic
field, $\beta$ is nominally infinite.  The initial resolution parameter
$Q_z$ has three maxima, each located where the vertical field crosses
the equator.  From inside out, these maxima are $Q_z = 15$, 29, and
20; the linear growth phase of the MRI should be well-resolved.

The fiducial model was run for $6.6\times 10^4 M$ in time, corresponding
to 54 orbits at the initial pressure maximum, and over 1000 orbits at
the location of the ISCO, $R=6M$.
This evolution time is longer than previous global simulations,
and permits an examination of the long term behavior of the accretion
flow and MRI-driven turbulence at these resolutions.  The evolution
proceeds in several stages as identified by the evolution of the total
magnetic energy (Fig.~\ref{fiducialenergy}).  First there is a period
of exponential growth in the magnetic field.  This is rather brief;
by $t=1000 M$ (about 1.3 orbits at $R=24M$), significant radial field
has already grown to supplement the vertical field in the innermost
part of the inner loop. This behavior is characteristic of the vertical
field MRI.  The total integrated radial field energy has doubled by this
time; the vertical field energy doubles by $t=1500 M$.  At first, the
toroidal field grows as the MRI creates radial field and the background
shear stretches radial field into toroidal; later azimuthal MRI modes
also contribute to toroidal field amplification.  By $t=2000M$ the
total toroidal field energy is 10 times as large as the total poloidal
field energy, and significant accretion (mass flow off the inner radial
boundary) has begun.  Global field energy continues to rise until it
peaks at $t\sim 8500 M$.  It then declines until $t=1.5\times 10^4 M$,
after which the volume-integrated energies of the different magnetic
field components vary slightly around a slowly declining trend line.
We will refer to the period between $2000M$ and $1.5\times 10^4 M$
as the ``initial peak"; the remainder of the simulation we call the
``quasi-steady state" period.

\subsubsection{Inflow equilibrium and the quasi-steady state}

Nearly every global disk simulation begins with a finite mass on the
grid.  Because accretion of some mass entails transfer of its
angular momentum to other mass, part of the matter in the simulation must
move outward as other mass moves inward.  Consequently, at most only
a portion of the disk can actually be in a state of inflow.  At best,
therefore, inflow equilibrium can be established only within some radius.

Having identified the largest radius within which inflow equilibrium can
be sought, one might define inflow equilibrium to be a state in which the
mass accretion rate is constant as a function of radius.  The problem
with this definition is that accretion is driven by the fundamentally
chaotic process of MHD turbulence.  The accretion rate at a specific
radius must therefore always be highly variable in time.  One solution
is time-averaging; regions of inflow equilibrium would then be those
ranges of radius over which the time-averaged accretion rate is constant.

A closely related procedure is to make use of the equation of mass
conservation
\begin{equation}
\frac{\partial \Sigma}{\partial t} + \frac{\partial \dot M}{\partial
R} = 0.
\end{equation}
Here $\Sigma$ is the surface mass density and $\dot M$ the accretion rate
integrated over the cylindrical surface at radius $R$.  Clearly, wherever
$\dot M$ is independent of $R$, $\Sigma$ is constant in time.  Thus, one
could also test $\Sigma(R,t)$ for time-steadiness across the radial range
of interest.  Equivalently, one could check that
$M(<R,t) = \int^R \, dR^\prime \, 2\pi R^\prime \Sigma(R^\prime,t)$
is time-steady.  This alternative has the conceptual advantage that it focuses
squarely on the primary matter of interest: that the mass distribution
in the disk does not change secularly over time.  It also has the technical
advantage that $\Sigma(R,t)$ changes more slowly than $\dot M(R,t)$, so
results taken from simulation data are less subject to noise fluctuations.

Figure~\ref{fig:massfillin} shows $M(<R,t)$ for the fiducial run.
The disk mass rises quickly during the first $10^4M$ in time and then
levels off.  After that time, there is a slow secular diminution in the
mass of the disk within $R=20M$, but its characteristic timescale is
quite long, $\simeq 5\times 10^4 M$ at $R=10M$, $\sim 8 \time 10^4 M$
at $R=20M$.  Because the time to drain and replenish the region inside
$R=20M$ is $M/\dot M \simeq 1.5\times 10^4 M$, this secular trend is
quite slow compared to the characteristic mass equilibration time.
Despite the overall steadiness of the radial mass profile, there are
also shorter timescale fluctuations that become progressively larger
in fractional terms at smaller radii, reaching $\sim 50\%$ at $R=10M$.
These reflect, of course, the continuing large amplitude fluctuations
in the mass accretion rate both as a function of time and of radius.
The slow diminution in inner disk mass reflects the declining trend
in the magnetic energy, which leads to a parallel fall in the mass
accretion rate.

Another way to test for inflow equilibrium is to compare the
simulation time with an estimate of the characteristic inflow time.
We can compute an average inflow velocity from simulation data by taking
\begin{equation}\label{eq:inflow1}
\langle v_R (R) \rangle = { \int \rho v_R R d\phi dz \over 
\int \rho R d\phi dz} .
\end{equation}
These values are averaged over time to remove the ever-present
fluctuations.  The accretion time from radius $R$ is then
$t_{in}(R) = \int^R dR^\prime / \langle v_R (R^\prime)\rangle$.
Computing this for the fiducial run at $R=20$ we obtain $1.2\times
10^4 M$, consistent with the estimate above based
on the average accretion rate and the disk mass interior to $R=20M$.

How does this result compare to an estimate obtained from steady-state
disk theory?  In the steady-state limit the equation of angular
momentum conservation is
\begin{equation}
W_{R\phi} = { \dot M \Omega \over 2\pi \Sigma} \left(1 - j_*/j\right)
\end {equation}
where $W_{R\phi}$ is the vertically averaged $R\phi$ component of the stress
tensor, and $j$ and $j_*$ are the specific angular momentum at $R$
and the angular momentum accreted per unit mass.  The $j_*$ term
determines the net flux of specific angular momentum; traditionally,
it has been set to the angular momentum at the ISCO on the assumption
that stresses cease there.  Following \cite{Shakura:1973}, we write the
vertically-integrated stress in units of the vertically-integrated pressure
and assume there is a single temperature throughout the flow; i.e., we set
$W_{R\phi} = \alpha \Sigma c_s^2$.  With those assumptions, we obtain
for the steady-state mean infall velocity
\begin{equation}\label{eq:inflow2}
\langle v_R \rangle_{SS} = {\dot M \over 2\pi R \Sigma}
= {\alpha c_s^2 \over R \Omega } \left(1 - j_*/j\right)^{-1}.
\end{equation}
Agreement between the inflow velocities from (\ref{eq:inflow1}) and
(\ref{eq:inflow2}) would indicate that the observed mass accretion is
in approximate steady state at a rate consistent with the angular
momentum transport produced by the observed stress.

We test this for the fiducial run for a time average over $10^4M$
beginning at $t=4\times 10^4 M$.  We compute a vertically and
time-averaged Maxwell stress, pressure, and density to obtain $\alpha$
and $c_s^2$.  The average value of $\alpha$ thus obtained for the Maxwell
stress is $0.017$.  To account for the Reynolds stress (not directly
measured in global simulations), we increase the value of $\alpha$ in
(\ref{eq:inflow2}) by 25\%, a value consistent with results from shearing
box simulations.  The velocity derived from the steady state disk model
(with $j_* = j_{\rm ISCO}$) and the velocity obtained directly from the
accretion rate and the mass are compared in Figure~\ref{fig:alphacomp}.
The agreement is good out to just beyond $30 M$.  This match is consistent
with the range over which the time-averaged $\dot M$ is constant with
radius, the computed $t_{in}$ at $30M$ which is $4.5\times 10^4 M$, and
our estimates based on the evolution of $M(<R,t)$ as described above.
The two curves deviate outside of $R\sim 32M$ because there the infall
time begins to exceed the simulation time.  At small radius the curves
deviate because the stress does not go to zero at the ISCO; reducing $j_*$
to $0.985 j_{ISCO}$ brings the curves into line right down to the ISCO.

\subsubsection{Comparison with shearing box results}

The initial transient and quasi-steady periods are seen in both global
and shearing box simulations.  Figure~\ref{fig:stratcomp} plots the total
magnetic field energy, normalized to its peak value, as a function of time
in units of orbits at the initial pressure maximum.  Overlaid on this
is the evolution of the total magnetic field energy in the stratified
shearing box simulation with 16 zones per $H$, again normalized to the
initial peak value.  Because the early evolution of the global simulation
is relatively local, dominated by MRI growth in the confined region where its
growth rate is greatest (i.e., the inner rings of the initial torus),
it is perhaps not surprising that the initial evolutions are similar.
More interesting, perhaps, is the similarity between the two
models during the subsequent quasi-steady state phase.  Between 20--50
orbits in each run, the ratios of the total average radial to toroidal
magnetic field energy, $\brsq$, are 0.068 and 0.070 for the shearing box
and the fiducial model, respectively.  This similarity suggests that the
effective resolution in the fiducial global simulation, at 14--60 zones
per $H$ depending on radial location, is comparable to the 16 zones per
$H$ used in the shearing box.  A comparison with Fig.~\ref{fig:shrstress}
shows that higher resolution shearing box simulations have a smaller
decline in stress immediately after the initial peak.  The 16 zone
shearing box simulation, however, sees a regrowth of field energy beyond
orbit 50 that is not seen in the global simulation.  The reason for
this is uncertain.  It could be due to the better azimuthal resolution
in the shearing box, possibly by better capturing a (nonaxisymmetric)
dynamo process.  Another possibility is that the observed field regrowth
is due to other properties associated with the shearing box, e.g.,
restricted box size, shearing-periodic boundaries, and overall symmetry.

Among the differences between shearing box and global simulations is the
latter's large dynamic range in $\Omega$, which is important because the
MRI $e$-folds at a rate $\sim \Omega$.   As a result, 
a single spatially-averaged value is not very useful for comparison
with shearing boxes.  Instead, we show
figures of the principal diagnostic quantities as a function of radius,
averaged in time and azimuthally- and vertically-averaged weighted
by density (Figs.~\ref{fig:twoloopdiags} and \ref{fig:phiresdiags}).
In each case, the averaging interval was chosen to be the longest
time-span covering the quasi-steady epoch for all simulations shown in
a given figure; consequently, the averaging periods for the different
figures are different.  Likewise in each case, the radial range was
restricted to the region defined as the ``inner disk,'' $6M < R < 20M$
in order to focus attention on the portion of the simulation most closely
resembling a statistically time-steady accretion flow.

Data from the fiducial run appear in both of these figures.
Its density-weighted $\Qz$ drops steadily inward, from $\simeq 8$ at
$R=20M$ to $<2$ at $R=6M$.  By contrast, its $\Qp$ value varies only
slightly with radius, remaining near $\simeq 8$--10 throughout the
inner disk.  Both behaviors---the strong dependence of $\Qz$ on radius
and the near-constancy in radius of $\Qp$---are characteristic of all the
simulations.  $\brsq$
displays a radial profile very similar to that of $\Qz$, falling from
$\simeq 0.08$ in the outer part of the inner disk to $\lesssim 0.04$
just outside the ISCO at $R=6M$.

Thus, even with a poloidal cell-count of $256 \times 288$, the fiducial
run is, at best, marginally resolved according to both the $Q_z$ and
$Q_\phi$ criteria.  There is adequate resolution to describe linear
MRI growth of poloidal perturbations only near $R=20M$; nowhere in the
inner disk is it well enough resolved to describe poloidal nonlinear
behavior properly.  The azimuthal resolution is no better: $\Qp$ never
reaches the $\simeq 20$ level indicated by shearing box simulations.
Similarly, the value of the nonlinear criterion $\brsq$ is at most only
about half what the shearing box simulations suggest is characteristic of
well-resolved turbulence.  It is near to the value seen in the marginally
resolved 16 zones per $H$ shearing box.

The radial gradient in $\langle Q_z\rangle$ is an illustration of how
grid-design can interact with physics.  The definition of this quantity
is $2\pi |v_{Az}|/(\Omega \Delta z)$.  In the cylindrical coordinates
used here, $\Delta z$ is independent of radius.  Consequently, $Q_z
\propto v_{Az}(R) R^{3/2}$; unless $v_{Az}$ rises rapidly inward, poloidal
resolution quality falls toward small radius.  Here the average $v_{Az}$
goes roughly like $R^{-1/2}$ in the inner disk so $\Qz \propto R$.
By contrast, $Q_\phi \propto v_{A\phi} R^{1/2}/\Delta\phi$, leading to
its much weaker dependence on $R$.  The end-result of these different
dependencies on radius is that, unless $Q_z$ is very large at
larger radii, the linear growth rate of axisymmetric MRI modes will be
reduced in the inner disk even while non-axisymmetric modes continue to
grow at their correct (albeit slower) rate.

The radial gradient in $\Qz$ also provides a finely-sampled
measure of convergence properties.  At each radial cell, these global
simulations sample a different effective resolution.  As we will discuss
in \S~\ref{sec:interp}, this effect allows us a much more quantitative
measurement of the convergence rate than we would otherwise be able
to obtain.

\subsection{Different grid resolutions}

Table~\ref{table:comphi} lists time- and volume-averaged values for various
parameters, both diagnostic and physical, as well as the time-averaged
accretion rate through the inner $R$ boundary as a fraction of the initial
torus mass.  The time averages are
taken over the steady state period for each simulation; the volume
averages are limited to the inner disk body.  For those parameters
with potentially significant radial gradients, a range of values is given:
the first number refers to the value at $R=6M$, the second to $R=20M$.
The scale height $H$ is defined by a time- and density-weighted mean
of $\sqrt{2}c_s/\Omega$.  In these simulations, in which total energy
is not conserved, $c_s$ scales with radius roughly $\propto R^{-1/2}$,
so that $H$ is approximately $\propto R$.  In most cases, $\alpha_{mag}$
increases outward, but simulation twoloop-1000-hr is an exception: in that
case, $\alpha_{mag}$ decreases slightly toward larger radius.

\subsubsection{Azimuthal Resolution}

We next consider how the results seen with the fiducial model change as
the resolution is altered.  We begin with a study of the $\phi$ resolution
(Fig.~\ref{fig:phiresdiags}).  Model twoloop-1000-mlp uses grid PL, which
decreases the number of $\phi$ zones to 32 while gaining a modest increase
in radial resolution due to bringing in the outer boundary somewhat.
Twoloop-1000-mhp uses grid PH, which is the same as the M grid, but with
an increase in the number of $\phi$ zones to 128.  As the orbital speed
generally sets the Courant limit in these simulations, this high-$\phi$
resolution model is more costly to evolve.

Figure~\ref{fig:comphi} shows the evolution of the total poloidal magnetic
energy as a function of time for the three runs.  For the 32 zone run,
exponential growth ends earlier compared to the other simulations, with
a more gradual climb to a peak value.  The 128 zone run leaves the rapid
growth stage earlier than the 64 zone run, but when it levels off after
declining from its magnetic energy peak, it does so at a higher level.
Thus, the sustained field energies after the initial peak are clearly separated
by resolution.

As shown in Figure~\ref{fig:phiresdiags}c, for
$R \gtrsim 12$--$15M$, $\Qp$ increases by roughly a factor of
2 for each factor of 2 improvement in resolution.  
Notably, however, both poloidal indicators, $\Qz$ and $\brsq$, also
respond positively to improvement in toroidal resolution at larger radii
within the inner disk.  Like $\Qp$, the response to improved resolution
is even stronger between the M and PH grids than between PL and M.
In other words, {\it improved azimuthal resolution
leads to stronger poloidal field, even in the absence of improved
poloidal resolution}, and this effect strengthens with finer azimuthal
resolution.

\begin{deluxetable}{l|cccccccc}
\rotate
\tabletypesize{\scriptsize}
\tablewidth{0pc}
\tablecaption{Resolution Comparisons\label{table:comphi}}
\tablehead{
 \colhead{Name}&
 \colhead{$R,\phi, z$ zones}&
 \colhead{$\dot M$}&
 \colhead{$\alpha_{mag}$} &
 \colhead{$Q_z$} &
 \colhead{$Q_\phi$} &
 \colhead{$B_R^2/B_\phi^2$} &
 \colhead{$B_{z}^2/B_{\phi}^2$} &
 \colhead{Cells per $H$}
}
\startdata
{\bf
twoloop-1000-mr} &$256\times 64\times 288$ & $2.71\times 10^{-6}$ &0.24--0.33& 1.6--7.5 & 7.9--10. & 0.040--0.077 & 0.0088--0.018 & 11--60\\
twoloop-100-mr   &$256\times 64\times 288$ & $2.51\times 10^{-6}$ & 0.29--0.37 & 3.0--9.0 & 12.--11. & 0.054--0.092 & 0.012--0.023 & 14--65\\
twoloop-1000-mlp &$256\times 32\times 288$ & $1.99\times 10^{-6}$ &0.25--0.24 & 4.2--4.5 & 8.5--6.0 & 0.059--0.046 & 0.016--0.012 & 8--49\\
twoloop-1000-mhp &$256\times 128\times 288$& $4.40\times 10^{-6}$ &0.23--0.41 & 2.1--11. & 22.--22. & 0.045--0.13 & 0.0079--0.035 & 13--57\\
twoloop-1000-lr  &$128\times 64\times 144$ & $0.94\times 10^{-6}$ &0.095--0.25& 0.46--2.3 & 5.7--9.4 & 0.019--0.050 & 0.0055--0.0081 & 8--30\\
twoloop-1000-hr  &$816\times 64\times 288$ & $1.89\times 10^{-6}$ &0.31--0.29 & 3.8--4.9 & 13.--11. & 0.064--0.057 & 0.021--0.022 & 10--46\\
oneloop-1000-mr  &$256\times 64\times 288$ & $3.85\times 10^{-6}$ &0.29--0.35& 2.8--9.1&12.--11. & 0.059--0.086 &0.012--0.024 & 14--60\\
oneloop-100-lr   &$128\times 64\times 144$ & $1.43\times 10^{-6}$ &0.13--0.29& 0.42--3.3 & 5.6--12. & 0.017--0.065 & 0.0053--0.013 & 9--36\\
oneloop-100-mr   &$256\times 64\times 288$ & $7.92\times 10^{-6}$ &0.30--0.38& 3.0--10. & 13.--12. & 0.063--0.10&0.010--0.033 & 13--76\\

\enddata
\end{deluxetable}

\subsubsection{Poloidal Resolution}

Figures~\ref{fig:oneloopdiags} and \ref{fig:twoloopdiags} show the
diagnostics for one-loop and two-loop initial field configurations with
a range of resolutions.  In the one-loop cases,
a factor of 2 improvement in poloidal resolution yields an equal factor
in $\brsq$ and a factor of 3 improvement in $\Qz$.  Moreover, even with
a grid of $256\times 64\times 288$ zones, these simulations remain unresolved.
The best $\Qz$ was only $\simeq 10$, while it fell
below 6 inside $R \simeq 10$; $\brsq$ hovered around 0.1 outside $R \simeq 15M$,
falling to 0.06--0.07 near the ISCO, whereas the
shearing box simulations indicate a converged value of almost 0.2.

Within the two-loop series of simulations, we find that the move from the
L poloidal grid to the M yielded the same level of improvement as in
the one-loop case: a factor of 2 in $\brsq$ and a factor of 3 in $\Qz$.
However, progress stalls when only the radial zone size is reduced,
as in the R grid.  Outside $R\simeq 10$--12, the
finer radial resolution actually led to a deterioration in both $\brsq$
and $\Qz$.  We speculate that this may be due to the highly-elongated
cells created in this grid by its finer radial resolution.   In
contrast to the M grid, in which $R\Delta\phi/\Delta R \simeq 0.4$--1.4,
that ratio is 4--6 in the R grid, increasing toward larger radii.

Although not illustrated in the figures, much the same can be said about
the effect of resolution on $\alpha_{mag}$.  When the magnetic geometry
is held fixed and only resolution changes, finer resolution in both the
poloidal and azimuthal grid lead to larger values of $\alpha_{mag}$.

Overall, it appears that improving radial resolution without also
improving either vertical or (perhaps especially) azimuthal resolution
does not yield significant progress toward convergence.  As we have seen,
the linear indicators ($Q_z$ and $Q_\phi$) do not grow (if anything,
they fell in this case), nor do nonlinear indicators such as $\brsq$
improve.

\subsection{Different initial magnetic fields}

All simulations must begin from some particular magnetic field.
The problem, of course, is that we can only guess at what might actually
occur in Nature.  We have no way to say what geometric structure it
might have, yet numerous variations are imaginable.  The field may or may
not have net flux; even if it has no net flux, it may be predominantly
toroidal or poloidal; and numerous sorts of poloidal configurations are
possible.  Even granted the field geometry, we must still specify its
initial strength; so long as it corresponds to a plasma $\beta \gg 1$,
we have no guidance in this respect either.  The most we can hope for is
that in the end the resulting steady state accretion flow is independent
of our arbitrary choices.

We begin with the question of sensitivity to initial field strength with
a comparison of oneloop-100-mr and oneloop-1000-mr.  Stronger fields,
of course, will have an initial advantage of larger $Q$ values at
a given resolution.  In the quasi-steady state period, as shown by
Figure~\ref{fig:oneloopdiags}, a factor of 10 in initial magnetic field
pressure makes a consistent, but very small difference: the stronger
initial field in general has larger values of the diagnostics, but only
by $\sim 10\%$.  There is a greater contrast between twoloop-100-mr and
twoloop-1000-mr, generally $\simeq 20$--30\% and in some places larger.
Like the one-loop case, the simulation with the initially stronger field
retains that advantage.

The main effect of a stronger initial field is in the initial evolution;
it creates stronger accretion early on and fills the inner disk with
matter at a faster rate.  This happens for two reasons, neither of which
has much long-term effect.  The first is that the stresses driving
the initial inflow are not created by correlated MHD turbulence, but
by the shear that transforms the initial poloidal field into toroidal.
The second is that in the initial state, $\Qz$ is smaller if the field
is weaker, depressing the linear MRI growth rate in some locations.
In the end how the mass
distribution reaches its steady-state is not important, so long as one
evolves long enough to reach the steady-state.

We next turn to the question of field geometry.  \cite{Beckwith:2008a}
demonstrated that the magnitude of the axial funnel field attached to the
black hole, capable of powering a jet when the black hole spins, depends
strongly on whether the vertical component of the initial magnetic field
in the simulation has a consistent sense over a wide span of radii---the
jet luminosity is substantial only when it does.  They also showed that
the accretion flows in the cases they considered depended much less on
the field structure, although the stress levels and accretion rates are
noticeably lower when the initial field is purely toroidal.  Here
we take a closer look at the accretion flows associated with two
specific initial field configurations: one initial dipole loop versus two.

The first point is that the two-loop configuration does not persist
throughout the simulation.  Distinguishing the loops by their sign of
$A_\phi$, the azimuthal component of the vector potential, we find that in
the fiducial model the disk inside $R=20M$ is effectively dominated by the
inner loop until $t \sim 10^4 M$.  By the end of that period, much of the
inner loop's flux has been accreted through the inner boundary, and the
remainder has risen to high altitude and left the disk proper.  For the
next $\sim 10^4 M$, the main body of the disk is entirely dominated by the
outer loop.  Between $\sim 2.2\times 10^4M$ and $\sim 3.2\times 10^4M$,
enough inner loop flux has settled into the disk that it is very mixed.
During this period, there is significant reconnection between oppositely
directed field loops.  For the remainder of the simulation (until $t =
6.6\times 10^4 M$), the disk is once again dominated by poloidal field
loops having the sense of the outer of the two initial loops.  In other
words, with the exception of an episode that lasts only $\sim 1/6$ of the
simulation, there is little to distinguish the poloidal field geometry in
the inner disk of a two-loop simulation from one that begin with only a
single set of poloidal loops.  In both cases, there is a consistent sign
of $A_\phi$ (i.e., consistent sense of field circulation) in the inner
disk, but with a highly complex structure (the more fully-developed the
turbulence, the more complex, of course).

Nonetheless, our time-averaged diagnostics do show interesting contrasts
between the one-loop and two-loop results (Figs.~\ref{fig:oneloopdiags} and
\ref{fig:twoloopdiags}).  At low-resolution, the $\Qz$ diagnostic improves
by $\sim 50\%$ from two-loop to one-loop at all radii.  At medium-resolution,
$\Qz$ in the one-loop case is only about 10\% greater in the disk body
(e.g., $R=20M$) than for the two-loop case, but that advantage widens toward
smaller radii, rising to almost a factor of 2 by $R=10M$.  Very similar
contrasts are found for $\brsq$.  Although the magnitude of the change
is smaller, the trend is the same for $\alpha_{mag}$.

One might reasonably ask why these contrasts occur, given our assertion
that the inner disk magnetic field at any single time generally
has a uniform sign of $A_\phi$. A clue comes from looking at their
time-variation.  During the period $1$--$2\times 10^4 M$, the diagnostic
quantities in twoloop-1000-mr are, in fact, very similar to those of
oneloop-1000-mr.  Their decline at small radii takes place only after
$2\times 10^4M$, the time when oppositely-directed field loops reenter
the disk.  That re-entry is extremely irregular and creates gradients
in $A_\phi$ on very short lengthscales.  For the same gridscale, the
larger amount of short lengthscale fluctuation power leads to a faster
numerical reconnection rate.  This, in turn, weakens the magnetic field
to the point that the MRI is unable to rebuild the field strength, even
after the period of enhanced reconnection is over.  Throughout this
later period $Q_z$ is only $\simeq 2$--$8$, insufficient to drive the
MRI at its full rate.

Thus, the two-loop initial condition creates---through
a rather recondite mechanism---considerable short length-scale power
that requires a finer resolution grid to treat properly.  In principle,
a weakened magnetic field (and therefore smaller values of $\Qz$ and
$\Qp$ for a fixed gridscale) might be a physical consequence.  However,
there is clear evidence that this really is a resolution effect.  As we
have just pointed out, the prevailing $Q_z$ after the strong reconnection
epoch is inadequate to drive linear MRI growth, much less support nonlinear
poloidal dynamics.  The low value of $\brsq$ further supports our conclusion
that the turbulence is less than fully-developed at late times in the
two-loop simulations.

%% file: interpretation.tex
\section{Interpretation}
\label{sec:interp}

The principal goal of this paper is to develop from highly resolved
shearing box models a set of diagnostics
that can be applied to gauge how well global simulations
represent the quasi-stationary behavior of a disk obeying the same
physics as in the simulation.  In this section we will discuss how to
use the diagnostics we have identified, making use of their application
to the simulations described in the previous section.

One might have supposed that the most natural quantity to study
as a gauge of progress toward convergence is the stress, the physical
quantity of central interest to the entire subject.  However, there are
several good reasons {\it not} to use it.  The first is that it is not
dimensionless.  A similar statement can be made about the accretion rate.
It can be useful to compare quantities such as stress and accretion
rate between a set of simulations with the same initial conditions, but a
comparison of any dimensional value across different simulations would
need calibration in terms of other quantities.

The stress is often quoted in units of the gas pressure, giving it an
apparently ``natural" dimensionless form, but it is not at all clear that
this form is appropriate.  We do not know whether the pressure is the only
variable influencing the stress, or whether the dependence is linear.
Fluctuation timing correlations \citep{Hirose:2009} suggest that the
relationship is more complicated than commonly thought.  In addition, even
if stress were dependent only on the pressure and exactly proportional
to it, the great majority of accretion simulations treat thermodynamics
with very crude approximations, making the value of the pressure both
questionable and difficult to compare from one simulation to the next.
For all these reasons, therefore, we prefer to use diagnostics that are
dimensionless ratios of quantities referring only to magnetic properties.

\subsection{Relationship between the different diagnostics}

Each of our diagnostics has a slightly different standing.  $\Qz$ and
$\langle Q_\phi\rangle$, for example, depend directly on cell sizes
in the grid and can be meaningfully evaluated on the initial data
whenever it includes either vertical magnetic field (for $\Qz$) or
toroidal field (for $\Qp$).  In that
sense, they are predictive measures of the ability of the simulation to
describe accurately the linear growth of, respectively, the axisymmetric
and non-axisymmetric branches of the MRI.  At the same time, however,
both are also simulation products because their values change as the
field strength evolves; in that sense, they also serve as diagnostics
for the fully developed MRI-driven turbulence.

The quantity $\brsq$ has little significance in the initial state;
it is simply a function of the initial magnetic field. It is
not even defined if the initial field is entirely poloidal.  Thus, it is
purely a product of the simulation; the same is true of $\alpha_{mag}$.

The definitions of the $Q$ parameters have an explicit dependence on
grid size; their values can never converge with increased resolution.
In contrast, $\brsq$ and $\alpha_{mag}$ measure physical quantities and
do not have an explicit dependence on resolution in their definitions.
In principle, they could be measured in a laboratory experiment.  Thus,
they signal progress of the simulation toward achieving the true physical
state of the MHD turbulence, no matter what resolution or algorithm has
been used.

Despite these contrasts in status, there is a tight relationship between
these different diagnostics of resolution quality.  This relationship is
illustrated in Figure~\ref{fig:qzvsbrby}.  Although the filled circles
in this figure are drawn from 11 different times at all radial cells
in the inner disk in each of 3 different simulations (twoloop-1000-lr,
twoloop-1000-mr, and twoloop-1000-mhp) that differ in both poloidal and
azimuthal griding, they closely follow a single track: $\brsq \simeq
0.01 ~{\Qz}^{0.65}$ for $\Qz \lesssim 10$--12, flattening out at larger
$\Qz$, where it rises slowly to almost 0.2.  The crosses, squares, and
diamond represent the time-averaged values in the midplane region of the
different stratified shearing box simulations; the different symbols
distinguish different ranges of $\Qy$ (the crosses have similar $\Qy$
to $\Qp$ in the global simulations; the others have larger $\Qy$).
The shearing box simulations with $\Qy$ similar to the $\Qp$ values
of the global simulations lie very close to the track defined by the
global simulations.  Consonant with our finding that better azimuthal
resolution also strengthens poloidal field, the shearing box results
with larger $\Qy$ lie somewhat above the track; that is, larger $\Qy$
can compensate to a certain degree for smaller $\Qz$.  Taken as a whole,
this figure confirms that for global simulations as well as for shearing
boxes, convergence in $\brsq$ does not begin until $\Qz \gtrsim 10$--15
for typical global simulation azimuthal resolution (i.e., $\Qp \simeq
10$); larger $\Qy$ ($ \gtrsim 25$) can relax this requirement slightly.

\subsection{Importance of $y/\phi$-resolution}

In most of the MRI literature hitherto, discussion of resolution criteria
has generally focused on poloidal griding, particularly the $Q_z$
criterion established by \cite{Sano:2004}.  Less attention has been
paid to the $\phi$ resolution.  It is generally understood that in
multi-dimensional simulations one should avoid cell aspect ratios that
are too far from unity.  Accuracy in directionally split algorithms
requires that the truncation errors in the different spatial directions
not be too different.  In any case, if the truncation error in one
spatial direction remains large due to a lack of resolution,
improvements in resolution in other dimensions will be of limited
value due to the dominance of that error.

In the context of disk dynamics, including shearing box simulations,
experience suggests that even though it is best for $\Delta R/\Delta
z \sim 1$, ratios $R\Delta\phi/\Delta R$ as large as $\sim 4$ are
acceptable because orbital shear tends to stretch out features in the
azimuthal direction.  Our findings here suggest that larger ratios, such
as are found in twoloop-beta1000-hr, tend to weaken the development of
MHD turbulence even though the unfavorable ratio was created by improved
radial resolution.

To our knowledge, the comparisons presented here are the first to
underline the importance of adequate $\Qp$.  Our analysis indicates that
sufficiently large $\Qp$ is necessary for the proper development of
poloidal properties such as $\brsq$ and maintenance of a large $\Qz$.
This is, perhaps, not so surprising as it is well-understood that
non-axisymmetric motions are essential for maintaining poloidal field
in the face of dissipation (the ``anti-dynamo'' theorem).  Even the
earliest stratified shearing box simulations found evidence for a dynamo
process \citep{Brandenburg:1995}, and recent simulations have shown
explicitly that magnetic field evolution can be empirically modeled with
a simple $\alpha$--$\Omega$ dynamo equation \citep{Guan:2011,Simon:2011}.
Although we have yet to develop a detailed understanding of the mechanism
by which the MRI creates the $\alpha$-effect that generates poloidal field
from toroidal, it clearly requires adequate resolution to be effective.
Future simulations should therefore be graded on their values of $\Qp$
as well as $\Qz$.

Finally, one should bear in mind that $\phi$ resolution might well
be much more important in global simulations compared to local.
The shearing box is in a local co-rotating frame whereas the global
simulation must deal with the full orbital velocity.  If one assumes
that the numerical dissipation level in a code is proportional to a
diffusion coefficient, e.g., $\Delta x^2/\Delta t$, then the diffusion will
be roughly proportional to the fastest velocity times $\Delta x$ because
$\Delta t \sim v_*/\Delta x$, where $v_*$ is that fastest velocity.  For a
global simulation, the orbital velocity is $R\Omega \propto (R/H)
c_s$.  Because for a shearing box the
fastest velocity is $\le c_s$, one might well require $R/H$ more
grid zones in the $\phi$ direction compared to a shearing box
in order to have similarly low levels of diffusion.
For global thin disk simulations this requirement
is particularly problematic.

\subsection{Connection with stress}

It would seem natural that stronger magnetic fields would lead to larger
stress.  In fact, the correspondence between the hierarchy we see in our
diagnostics ($\Qz$, $\langle Q_\phi\rangle$, $\brsq$, $\alpha_{mag}$)
is reproduced closely in stress levels.  Figure~\ref{fig:stressprofcomp}
plots the stress as a function of radius and can be compared with the
previous figures showing diagnostic quantities versus radius.  As the
diagnostics indicate better resolution, the stress rises.  Moreover, when
poloidal resolution is improved and the general level of stress rises, so
does the ratio between stress near the ISCO and stress in the disk body.
Although not illustrated in the figure, this trend also applies to the
radial stress profile at particular times during an individual simulation.
Higher stress levels at larger radius generally correspond to higher
stress throughout the flow, even if the stress and the $Q$ values decrease
moving in.  The ISCO-region stress appears even more sensitive to the
prevailing stress than the stress in other regions of the inner disk.

Because even relatively low resolution simulations nevertheless show
some stress due to MRI-turbulence, the mere presence of stress and
accretion are not themselves indicative of convergence in any sense.
Weak fields remain unstable at longer (better resolved) wavelengths,
but at greatly reduced growth rates.  As we have tried to make clear,
not even the best-resolved global simulations shown here appear to be near
the diagnostic levels identified as adequate in the shearing box models.
Even if some regions satisfy the resolution criteria, their variation
with radius means they are not satisfied {\it globally}.  We expect that
still better-resolved simulations can be expected to show both higher
stress levels overall, and a greater ratio of ISCO-region stress to
stress in the disk body.

\subsection{Effects of the initial magnetic configuration}

Because the MRI grows exponentially when the initial field is weak
($\beta \gg 1$), it has long been hoped that the saturated state of this
instability would be independent of the initial field intensity.  The
stress in unstratified shearing box simulations depends on the net value
of the initial field \citep[e.g.,][]{Hawley:1995b,Sano:2004}, but this
probably reflects the constraints imposed by the shearing box boundary
conditions, which preserve net magnetic flux.  In less-constrained global
simulations one might expect the field to evolve to some self-consistent
average magnetic energy with $\beta$ somewhere below equipartition,
independent of the field's initial strength, even though local regions
may have persistent net vertical flux \citep{Sorathia:2010}.  Although we
have not made extensive comparisons of simulations with different initial
values of $\beta$, those we have examined are consistent with this hope.
In the quasi-steady state period there are only modest differences between
our 1- and 2-loop simulations with initial $\beta=100$ and $\beta=1000$,
for example.

Field geometry can matter, however, in the way in which it interacts
with grid resolution.  Intuitively, one might suppose that a correct
discretized description of any field requires a grid with a characteristic
scale significantly smaller than the field's characteristic lengthscale
of variation.  Applying that argument in the present context suggests
that initial fields with more small-scale structure require finer
grids to describe accurately.  The contrast between our results for the
1-loop and 2-loop simulations supports this suggestion, although, as
we have explained, the specific way in which the small-scale structure
is created is rather more complicated than one might have guessed.
Introducing oppositely-directed field-loops creates more opportunities
for reconnection, and this rate is enhanced by coarse resolution.
Because the ability of a given grid to support field growth through
MRI stirring depends on the field strength itself when the resolution
is marginal, excessive numerical reconnection can lead to a long-term
suppression of the field intensity.

This has implications for comparing physically significant quantities
such as accretion rate or stress in simulations with differing magnetic
field geometry.  Without clear proof of convergence, the contrast between
them as simulated with the same grid is as likely to be due to a contrast
in {\it effective} resolution as it is to be due to a contrast in the
real physical behavior of the different systems.

\subsection{Inflow equilibrium}
\label{inflow}

In shearing boxes one measures the properties of the MHD turbulence after
many tens of orbits, when a quasi-steady state is clearly established.
In global simulations one looks for inflow equilibrium, a desirable
property because it is the primary prerequisite for achieving a
statistically stationary state.  Any other state could, in principle,
exhibit properties resulting more from transients due to arbitrary choices
in the initial conditions than to the intrinsic physics of the system.
Its meaning, however, requires precise specification in order to test
how well a simulation matches this condition.  The key question is
how long to evolve a global simulation in order to reach a reasonable
approximation to equilibrium.

We have used several procedures for determining inflow equilibrium.
One is to measure $M(<R,t)$, the mass interior to radius $R$ as a function
of time.  A related procedure is to compute the characteristic inflow time
from accretion rate and mass distribution, $M/\dot M$, and the average
inflow velocity $\langle v_R\rangle$.  This, in turn, can be compared
to the inflow velocity derived using the time-averaged stress and steady
state disk theory.  Agreement indicates that the observed mass accretion
rate is consistent with the observed angular momentum transport.  For the
fiducial run we find that all these measures give comparable results
and show that the disk inside of $R\sim 30 M$ is in inflow equilibrium.

It is worth noting that \cite{Penna:2010} proposed an alternative
definition of inflow equilibrium time at radius $R$, specifically $2
R/\langle v_R\rangle$ (rather than $\int dR/ \langle v_R \rangle$), solely
in terms of a general scaling argument.  Because the magnitude of $\langle
v_R \rangle$ increases rapidly inward, there is a large difference between
the local infall estimate $R/v_R$ and the actual integrated infall time.
The Penna et~al. value of the inflow equilibrium time is, for example, a
factor of 7 larger at $R=20M$ than the actual inflow time $\int dR/\langle
v_R\rangle$ in the fiducial run.  In fact, the inflow time estimated from
$2 R/\langle v_R\rangle$ at the initial torus pressure maximum exceeds the
time required to accrete the {\it entire} initial torus at the observed
$\dot M$.   

More generally, although an order of magnitude scaling argument may
provide a rough rule-of-thumb for estimating the equilibration time, it
does {\it not} provide a rigorous criterion.  Only observed properties of
the inflow can be used to test whether it is in equilibrium.  After all,
how a disk achieves a quasi-steady state is irrelevant; if the initial
condition (miraculously!) agreed identically with the steady-state
mass profile, it would be as good an equilibrium as one achieved after
long computation.  In fact, a commonly-used initial condition (one in
which the magnetic field is dominated by large dipole loops; our 1-loop
simulations are examples) behaves roughly in this way.  Large turbulent
magnetic stresses develop early in the simulation at the inner edge of
the initial torus, pulling mass into the inner disk far more quickly
than would happen at later, more representative, stages of evolution.
Once the inner disk is filled with matter, however, it evolves toward
an appropriate time-steady condition.  The large transient magnetic
stresses are simply a ``jump-start" that allows a more rapid approach
to the steady-state.

As a final comment on this topic, we also wish to emphasize that
achievement of inflow equilibrium does {\it not}, on its own, signal
numerical convergence.  In almost all of the global simulations we report
here, the late-time state was one of approximate steady-state in terms
of mass inflow within the inner disk, yet for none of them would we
claim numerical convergence.

\subsection{Resolution and Numerical Dissipation Rate}

In simulations such as these, in which there is no explicit resistivity
or shear viscosity, dissipation of turbulence occurs entirely due to
gridscale discretization error (and any artificial viscosity).  This is
true regardless of whether one uses an internal energy algorithm such
as {\it Zeus} or an energy conservative scheme such as {\it Athena}.
It follows that the numerical dissipation rate is affected by cell-size.
Because short lengthscale dissipation is the primary means by which
magnetic (and kinetic) energy is lost, the saturation strength of the
turbulence depends directly on this dissipation rate.  To understand
how resolution affects the amplitude of the magnetic field therefore
requires a closer consideration of gridscale losses.

In the shearing box simulations of \cite{Hirose:2006}, the
local rate of this sort of dissipation scales $\propto |{\bf J}|^{1.1}$
for electric current density ${\bf J}$.  In other words, by Amp\'ere's
Law, numerical dissipation is roughly proportional to the cell-by-cell
contrast in $|{\bf B}|$.  If the local scaling were {\it exactly}
proportional to $|{\bf J}|$, then the box-integrated dissipation would
be likewise proportional to the box-integrated $|{\bf J}|$.  However, if
there is any departure from linearity, it is necessary to be more careful
in regard to the distribution function of current density over the box.

The best data we have available for this purpose comes from the
simulations of Shi et~al. because their code explicitly tracked
the several different kinds of gridscale dissipation.  With remarkably
little scatter, this data shows that the box-integrated magnetic
dissipation rate $Q_{\rm tot} \propto J_{\rm tot}^{2.5}$, where
$J_{\rm tot}$ is the box-integrated current density.  Because this is
{\it numerical} dissipation, not physical, it must depend on the cell-to-cell
contrast in ${\bf B}$, rather than on the actual current.  Its most
appropriate dimensional form is therefore
\begin{equation}\label{eq:qtot}
Q_{\rm tot} = q \frac{\langle \delta B_{\rm grid} \rangle^{5/2}}{\Delta t}
              p^{-1/4},
\end{equation}
where $q$ is a number of order unity, $\Delta t$ is the time-step in
the simulation, and $p$ is the pressure.
We evaluate $\langle \delta B_{\rm grid}\rangle$ as
$\langle |\nabla^\prime \times {\bf B}|\rangle$,
where $\nabla^\prime$ denotes a {\bf curl} operator defined with respect
to cell-index rather than distance.  As resolution improves, one would
expect the cell-to-cell contrast to diminish, reducing $Q_{\rm tot}$.
However, there are countervailing effects: in turbulence, new short
lengthscale power can readily be generated by nonlinear transfer from
longer wavelengths; and the time-step $\Delta t$ also decreases.

In fact, the time-step $\Delta t$ depends on spatial resolution in two
ways: $\Delta t = \Delta x/c_s$, and $c_s$ depends on resolution because
Shi et~al. found that over their range of resolutions the total, i.e.,
gas plus radiation, pressure had not yet converged; it scaled roughly
$\propto (\Delta x)^k$, with $k \simeq -0.45$.  At fixed surface density,
hydrostatic balance implies that $c_s \propto p$, so $\Delta t \propto
(\Delta x)^{1-k}$.  Thus, we expect that the overall resolution scaling of
$Q_{\rm tot}$ should be $\propto \langle \delta B_{\rm grid} \rangle^{5/2}
(\Delta x)^{-4/3}$.  Comparing the $Q_{\rm tot}$ versus $\langle \delta
B_{\rm grid} \rangle$ relations found in the Shi et~al. simulations, we
find that the actual exponent of $\Delta x$ is $\simeq -1.3$, in very good
agreement with the dimensional argument leading to
Equation~(\ref{eq:qtot}).

When the magnetic field achieves its saturation strength, $Q_{\rm tot}$
balances the fluctuation power delivered to the grid scale by nonlinear
coupling with longer lengthscales in the MHD turbulence.  At resolutions
currently feasible, these couplings are non-local in the sense that the
shortest wavelengths couple significantly to essentially all longer
lengthscales; however, there is reason to think that in genuine MHD
turbulence, where the dynamic range between stirring scale and dissipation
scale is far greater, the dissipation scale actually decouples
\citep{Lesur:2010}.  Because all our simulations are in the regime in
which all lengthscales couple directly to the grid scale, we
write the rate at which longer scales couple to the dissipation scale as
\begin{equation}
H_{\rm nonlin} = s \Omega \langle \delta B_{\rm tot}^2\rangle \beta^{-m},
\end{equation}
where $\langle \delta B_{\rm tot}^2 \rangle$ is the total variance in the
magnetic field.  Nonlinear energy transfer occurs on the eddy turnover,
i.e., dynamical, timescale and involves the entire fluctuating power.
Equating $H_{\rm nonlin}$ to $Q_{\rm tot}$ yields the prediction
\begin{equation}
\delta B_{\rm tot} = \left[(8\pi)^m \frac{\Gamma q}{s}
         \frac{p_0^{m+3/4}}{\Sigma\Omega^2\Delta x_0}
         \frac{f_{\rm grid}^{5/2}}{(\Delta x/\Delta x_0)^{m+1-3k/4}}
         \right]^{1/(2m-1/2)}.
\end{equation}
Here we have introduced the adiabatic index $\Gamma$, column density
$\Sigma$, orbital frequency $\Omega$, characteristic pressure $p_0$,
and grid resolution $\Delta x_0$ for which $p=p_0$.  We have also introduced
the quantity $f_{\rm grid}$, the ratio between the characteristic amplitude
of magnetic field fluctuations on the grid scale and on the stirring scale.

As the cell-size $\Delta x$ decreases, we can expect $f_{\rm grid}$
to decrease.  However, only when convergence is reached must it scale
$\propto \Delta x^{m+1-3k/4}$.  Consequently, convergence in the magnetic
field intensity may not be monotonic.  Using the data in our stratified
shearing box calculations, we can measure how far toward convergence we
have progressed in this regard by computing $f_{\rm grid}$.  Combining the
Simon et~al. and Davis et~al. data, we find that $f_{\rm grid}$ falls
from 0.42 for the simulation with 8 cells per scale height to 0.26 when
there are 128 cells per $H$.  In other words, for these simulations, in
which $k=0$ and the resolution improves by a factor of 16, $f_{\rm grid}$
falls by only 40\% so that, if the index $m=0$, the ratio $f^{5/2}_{\rm
grid}/(\Delta x/\Delta x_0)^{m+1-3k/4}$ would rise by almost a factor
of 5 and the saturated field strength would fall by a factor of 25.
That the saturated magnetic field in fact increases, but rather slowly,
over this range of resolutions suggests that $-1 \lesssim m \lesssim
-0.5$; in other words, the nonlinear energy transfer rate increases as
the gas pressure rises relative to the magnetic pressure.  With this
dependence of the nonlinear energy transfer rate on the plasma $\beta$,
the saturated field strength increases as the gridscale fluctuations
become smaller relative to the total variance.  At still finer resolution,
the scaling with $\Delta x$ of both $f_{\rm grid}$ and the nonlinear
coupling could change, but at least in this range of resolutions for
these algorithms, the field intensity can be expected to grow with
further refinement in spatial grids.

Thus, the action of gridscale dissipation creates a direct relation
between resolution and the nonlinear processes that determine the magnetic
field intensity and Maxwell stress.  Because improving resolution also
increases short lengthscale fluctuation power, the level of gridscale
fluctuations diminishes only slowly with improving resolution.  If, as
appears to be the case in the simulations we have studied, nonlinear
coupling between large scales and small grows modestly with an increasing
ratio of gas pressure to magnetic pressure, the net result is a slow
increase in saturated magnetic intensity with improving resolution.
Processes like these account, at least in part, for the more stringent
requirements for resolution that we have uncovered.

\subsection{Difficulties of convergence testing}

Although convergence testing is an accepted approach for gauging the
reliability of numerical simulations, there are several issues that make it
extremely difficult to apply to global accretion simulations.  Some of these
difficulties are practical; others are more fundamental.

One of the fundamental difficulties arises because accretion is driven
by MHD turbulence, and the interaction of dynamics at widely-differing
lengthscales is intrinsic to the nature of turbulence.  Large dynamic
range in spatial scales generically demands very high resolution,
of course.  The difficulties associated with achieving convergence are
especially compounded for MRI-driven MHD turbulence for which stirring
occurs over a wide range of scales.  \cite{Lesur:2010} argue that a
clearly defined inertial range separating the stirring scale from the
dissipation scale still has not emerged in the shearing box, even at
the highest resolution available to their spectral code.  Because the
total Maxwell stress is dominated by field energy on the largest scales,
one may hope that its magnitude becomes reasonably well-determined
even while the resolution remains insufficient to describe details of
the short-wavelength dissipation, but that remains to be demonstrated.
The previous subsection discussed some of the problems to overcome in
order to achieve that decoupling.

Among the practical considerations is that MHD turbulence-driven global
accretion simulations must be three dimensional.  Achieving even modest
resolution in three dimensions requires a large number of cells.
Whenever the science pushes computational capacity to its limit, it
is infeasible to carry out additional simulations at higher resolution
than the level at which one begins to see interesting effects.  For this
reason, attempts are sometimes made to ``test convergence" by contrasting
one set of results with those found at coarser resolution.  Although this
may be the only available procedure for simulations carried out at the
largest feasible resolution, it is an approach that must be used with
caution.  If the gridscale at {\it every} resolution tested remains too
coarse to be sensitive to a physical mechanism, the results will give
the appearance of convergence, even while missing the very existence
of a mechanism that may be very important.  This motivates this paper:
using the better-resolved shearing box simulations to develop diagnostics
that allow a global simulation to be evaluated directly.

Finally, there is the difficulty of measuring the effects of truncation
error and the rate at which it is being reduced for simulations
carried out at relatively coarse, but practical, numerical resolutions.
Strictly speaking, every algorithm has its own properties, particularly
with respect to truncation error.  Truncation error is characterized
by the amplitude of the error, the effective convergence rate at a
given resolution, and the asymptotic convergence rate.  The latter
is a fundamental property of the method, but in practice one almost
never reaches the asymptotic regime, except in select test problems
or for one-dimensional problems where exceptionally fine resolution
may be achieved.   At resolution levels reached in typical
three-dimensional simulations, the observed convergence rate need not
agree with the asymptotic convergence rate of the algorithm.  Also,
the convergence rate alone provides no information about the amplitude
of the truncation error at a given resolution.  Again, by determining
properties that are characteristic of well-developed MRI-driven MHD
turbulence (e.g., $\brsq$), we sidestep this difficulty.

Both global and local simulations have made use of several different
algorithms (finite-difference: {\it Zeus}, GRMHD; finite-volume
conservative: {\it Athena}, HARM3D), as well as variations with in the
same general algorithm (e.g., different flux solvers).  The relative
merits of different algorithms will be reflected in the value of the
$Q$ parameters required to produce good results.  For the shearing box
results in \S~\ref{sec:shearing}, we found that the use of either {\it
Zeus} or {\it Athena} appears to make relatively little difference; our
diagnostics of convergence depend upon resolution in very similar ways.
In \S~\ref{conclusions} we compute the number of grid
zones that might be required for a future well-resolved global simulation.
These numbers can vary for one algorithm compared to another, according
to the $Q$ values that prove necessary.

\subsection{Spatial resolution requirements and published global simulations}

There is no single number by which spatial resolution may be defined.
Even to specify the raw cell count requires three numbers, one for
each dimension, and that is not sufficient because cell size depends
on the extent of the simulation volume.  Moreover,
what really matters is not so much the absolute size of the cells
as their size relative to the natural lengthscales of the problem.
For global simulations, the radius $R$ is the primary scale relevant
to the radial direction, so the appropriate measure of cell size is
$\Delta R/R$.  In the azimuthal direction, there are two natural scales:
radians and $2\pi v_{A\phi}/\Omega$.  The former is due to the necessity
of resolving a wide-range of non-axisymmetric modes in the turbulence;
the latter has to do with adequate resolution for the
fastest-growing linear non-axisymmetric modes.  In the vertical direction,
there are likewise two scales: the density (or pressure) scale height,
$H$, and $2\pi v_{Az}/\Omega$.  Quantities based on the Alfv\'en speed
always require time-averaging because the field evolves rapidly over the
course of the simulation; depending on the treatment of thermodynamics in
the simulation, the vertical scale height may also require averaging.
Therefore, to align resolution metrics against simulations in the
literature requires translation of all cell-size specifications into
these five natural units.

We have done this for a selection of published simulations in
Table~\ref{tab:cellsizes}.  In addition to cell size data, we list
the radius of the pressure maximum in the initial torus used, and
the simulation length in terms of orbits at that radius.  The data
from the various simulations reported here with the L, M, and R grids
are also given in this table.  The additional simulations used are:
the high-resolution poloidal-field simulation of \cite{Hawley:2002},
which we label HK02; the high-resolution Schwarzschild simulation of
\cite{Devilliers:2003b}, called KD0b; the quadrupolar field simulation
from \cite{Beckwith:2008b}, designated QD0; the high-resolution
thinnest simulation of \cite{Noble:2010}, which they labeled ThinHR;
the Schwarzschild simulation reported by \cite{Shafee:2008}, identified
as ``Shafee" in the table; and the fiducial simulation A0HR07
from \cite{Penna:2010}, here called ``Penna".  For many of these,
the published paper did not report full information.  For those for
which we have access to the data, we have computed what we needed;
for the remainder, we have filled in only those entries derivable from
the publications.  Where this required reading values off graphs, we have
added a question mark.  All but one of the simulations listed in the table
used an azimuthal range of $\pi/2$; the Shafee simulation used a wedge
only $\pi/4$ in angle.  In these other simulations, unlike the {\it Zeus}
simulations described in this paper, the radial grid has a fixed $\Delta
R/R$.  They also differ in that, unlike our new {\it Zeus} simulations,
which used cylindrical coordinates, they used spherical coordinates.
Because $\Delta z = R\Delta\theta$ in spherical coordinates, the radial
gradients in $\Qz$ seen in the {\it Zeus} simulations are greatly reduced.

\begin{deluxetable}{l|ccccccc}
\tabletypesize{\scriptsize}
\tablewidth{0pc}
\tablecaption{Cell-sizes\label{tab:cellsizes}}
\tablehead{
   \colhead{Name}&
   \colhead{$\Delta R/R$}&
   \colhead{$\Delta\phi$}&
   \colhead{$\Qp$}&
   \colhead{$\Delta z/H$}&
   \colhead{$\Qz$}&
   \colhead{$R_{max}$}&
   \colhead{Orbits}
}
\startdata
L Grid & 0.033--0.111  & 0.025 & 7--10      & 0.0333-0.1  & 1--3  & 35 & 33\\
M Grid & 0.017--0.055  & 0.025 &$\simeq 10$ & 0.016--0.08 & 2--10 & 35 & 54\\
R Grid & 0.004--0.0067 & 0.025 & 6--12      & 0.02--0.1   & 3--10 & 35 & 28\\
HK02   & 0.008--0.015  & 0.025 &            & 0.02--0.07  &       & 20 & 15\\
KD0b   & 0.024         & 0.025 & 19         & 0.06--0.14  & 6     & 25 & 14\\
QD0    & 0.024         & 0.025 & 26         & 0.08        & 8     & 25 & 14\\
ThinHR & 0.004         & 0.025 & 18         & 0.0086      & 25    & 35 & 12 \\
Shafee & 0.0065        & 0.025 &            & 0.03        &       & 35 &  8\\
Penna  & 0.013         & 0.05  &            & 0.064       & $\sim 5$? & 35 & 22\\
\enddata
\end{deluxetable}

As the data in this table demonstrate, most global simulations performed
hitherto have had effective resolutions roughly at the level of our M
grid with the standard azimuthal resolution, while the Penna simulation
most closely resembled our twoloop-1000-mlp simulation, which used our
poloidal M grid but coarser azimuthal resolution.  As a fraction of the
local radius, the radial cell sizes have been generally in the range
$\simeq 0.01$--0.025, similar to the 0.017--0.055 range of our M grid;
the only exceptions in this list are Shafee ($\Delta R/R = 0.0065$)
and ThinHR ($\Delta R/R = 0.004$), both of whose radial resolution was
closer to our R grid.  In terms of the scale height, the vertical cells
of most of these simulations are in the range 0.02--0.14, again matching
our M grid; by this measure, Shafee is similar to all the others, while
ThinHR is the only one substantially better, with $\Delta z/H = 0.0086$.
These values match the range explored in the stratified shearing box
simulations, where the coarsest we discussed had a vertical cell $0.125H$
thick, and the thickness of the finest vertical cell was $0.0078H$.
In radian measure, all of these simulations except Penna used a cell
with $\Delta\phi = 0.025$, identical to that in our standard M grid;
Penna's azimuthal cell was twice as large.  Our ability to be quantitative
about $\Qz$ and $\Qp$ is limited: we kept insufficient data from HK02
to compute these diagnostics, and we lack access to data from Shafee
and Penna.  However, on the basis of what the published papers say,
it is clear that $\Qz$ in most of the global disk literature has been
$\simeq 5$--8, while $\Qp$ has been $\simeq 20$.  By these measures,
at $\simeq 25$, ThinHR was a lone standout in terms of $\Qz$, but in
the same range as all the others in terms of $\Qp$.

Because our M and R grid simulations at best went only part way toward
convergence as measured by any of our diagnostics, we expect that all
these simulations likewise fell short of convergence.  For those cases
in which the initial magnetic field was in a single dipole loop (HK02,
KD0b), the $\Qz$ values suggest that the $\brsq$ measure is in the
neighborhood of $\sim 1/2$ its saturated value.  For those in which the
initial magnetic field was in two loops (QD0b, Shafee), the shift we have
seen in 2-loop simulations relative to 1-loop suggests that $\brsq$ is
somewhat smaller.  The Shafee simulation also has large aspect ratio cells
(although $\Delta R/(R\Delta\phi)$ is similar in ThinHR, the latter's
much finer vertical resolution appears to compensate).  In Penna, the
combination of a four-loop configuration and coarser azimuthal resolution
seems particularly challenging.  In our simulations such conditions
cause $\brsq$ to fall by a factor of 2 or more.  Such a reduction in the
relative magnitude of the radial component of the magnetic field would
indicate that the turbulence is not fully developed, that the field
intensity is rather lower than the converged value, and that the stress
is smaller than in saturation because it is proportional to $B_R$.

%% file: conclusions.tex
\section{Conclusions}
\label{conclusions}

Although global disk simulations have explored many important aspects of
the accretion process, their quantitative reliability remains uncertain.
In this paper we have made use of high resolution local shearing box
simulations to develop four diagnostics by which one may gauge how
closely a given simulation approaches to fully developed MRI-driven MHD
turbulence.  We have then examined how those diagnostics carry over to
global simulations.  Establishing this connection is an important step
in relating local simulations to global.  Local simulations will always
be able to include more physics and use better effective resolution
than global, and by means of this mutual calibration, information from
shearing boxes can help guide and interpret global models.

These four are: $\Qz$, the number of vertical cells across a wavelength
of the fastest-growing poloidal field MRI mode; $\Qy$ (or $\Qp$ in global
simulations), the number of azimuthal cells across a wavelength of the
fastest-growing toroidal field MRI mode; $\brsq$, the ratio of energy
in the radial magnetic field component to the toroidal component; and
$\alpha_{mag}$, the ratio of the Maxwell stress to the magnetic pressure.
Only the first has seen significant use previously, and we extend its
utility to gauge spatial resolution for nonlinear behavior of the MHD
turbulence as well as linear growth of the MRI.  Whereas \cite{Sano:2004}
found that a minimum $\Qz$ of 6--8 is required in order to describe
poloidal MRI linear growth, we find that the prerequisite for simulating
nonlinear behavior is more stringent and couples poloidal resolution
to azimuthal.  If the analogous azimuthal diagnostic $\Qp \gtrsim 20$,
then $\Qz \gtrsim 10$ is necessary; if $\Qp$ is smaller than that,
still larger values of $\Qz$ are required.

Using these diagnostics, we find that all the global simulations done
to date are under-resolved, particularly in the $\phi$ direction.
Only one simulation in the literature \citep[ThinHR,][]{Noble:2010}
comes close to adequate poloidal resolution, but even it does not meet
the azimuthal standard.  Our tests varying the azimuthal cell-size showed
that it can be important to the development of the poloidal, as well as
the toroidal, magnetic field.  Achieving adequate azimuthal resolution
is made especially difficult when the disk is thin for two reasons.
Thinner disks require smaller vertical cell dimensions, but avoiding the
deleterious effects of large cell aspect ratios then demands cells still
smaller in the azimuthal direction.  In addition, $v_A$ tends to diminish
with the smaller sound speeds seen in thinner disks; smaller $\Delta\phi$
is then necessary in order to achieve an adequate value of $Q_\phi$.

Regarding the initial conditions, we have found that the additional short
lengthscale fluctuation power that is a concomitant of more complicated
magnetic field geometry places stronger demands on spatial resolution.
The additional magnetic reconnection associated with attempting to
describe a more complicated field geometry on a given grid can weaken
the field sufficiently that MRI stirring is curtailed and the field
intensity remains artificially low.  Until global simulations are
adequately resolved it will be difficult to distinguish numerical from
physical effects arising from different initial field geometries.

We can make an estimate of the resolution required for a global torus
evolution using Equations (\ref{eqn:qz}) and (\ref{eqn:qy}).  To avoid the
problem of decreasing $H$ with decreasing $R$ that is found in cylindrical
coordinates, we assume a spherical-polar grid $(r,\theta,\phi)$.  To keep
$\Delta r = r \Delta \phi$ we use logarithmic spacing in $r$. We also
assume the radial grid spans a factor of 100 from the inner to the outer
boundary, and that the $\phi$ grid spans only $\pi/2$.  We next assume
$\beta = 10$, $\beta_z/\beta = 50$, $H/R = 0.1$.  Our target values are
$\Qz = 10$ and $\Qp = 25$ (one's preferred target $Q$ values may well
be algorithm dependent).  With these assumptions, the $\theta$ grid,
if equally spaced, must have $\sim 900$ zones if equally spaced from 0
to $\pi$.  In a practical simulation, however, one could increase $\Delta
\theta$ away from the midplane, reducing the number of zones required,
perhaps by as much as 50\% (450 zones).  The $\phi$ grid requires
200 zones and the $r$ grid 600 zones.  For $H/R = 0.1$, the number of
cells $(600\times 200 \times 450)$ is not too much greater than what has
already been used.  Of course, reducing $H/R$ would further increase the
number of required zones proportionally in each of the three dimensions.
Higher resolution also reduces the Courant-limited timestep, making it
challenging to evolve the disk for a large number of orbits.  In this
estimate we assumed a smooth distribution of field; intermittency in
the field distribution may change the required resolution from this
estimate. Typically there is significant spatial variation in $\beta$ so
that a density weighted $Q$ value can be larger than a volume-weighted
value.  On  the other hand, complex initial field distributions may
require more zones for adequate representation.  Future simulations done
at resolutions approaching this estimate can test these ideas.

Although we find that certain values of the resolution parameters indicate
well-resolved turbulence, lower values of those parameters don't mean
that turbulence necessarily decays rapidly; as we have described, it can
persist for long periods of time with lower amplitude and reduced stress.
Even under-resolved global simulations can have nonzero stress levels and
accretion rates.  A seemingly long duration of accretion is therefore {\it
not} a guarantor of convergence.  Qualitative conclusions can usefully
be---and have been---obtained from such simulations, but quantitative
stress levels are likely to be undervalued.

Lastly, we have explored the question of the meaning of ``inflow
equilibrium" in the context of global simulations.  Analysis of several
diagnostics can determine the existence and extent of the region of the
accretion disk that is in inflow equilibrium.  One can show that there
are at most weak long-term trends in the radial mass distribution, that
the time-averaged accretion rate is relatively constant as a function
of $R$, and that the observed accretion rate is consistent with the
observed angular momentum transport as computed from the steady-state
disk equation.  Even when the disk is in time-averaged inflow equilibrium,
there can be large short-term fluctuations in the mass distribution.
Moreover, whenever there is only a finite mass reservoir in the initial
condition, a sufficiently long simulation cannot, by definition, support
even a statistical steady-state for longer than the inflow time from
the initial half-mass radius.

In summary, this work provides guidance for future global simulations, 
both in terms of resolution and evolution time, to approach what is needed
for quantitative conclusions about accretion disk dynamics and structure.

%% file: figures.tex
\begin{figure}
\leavevmode
\begin{center}
\includegraphics[width=0.5\textwidth]{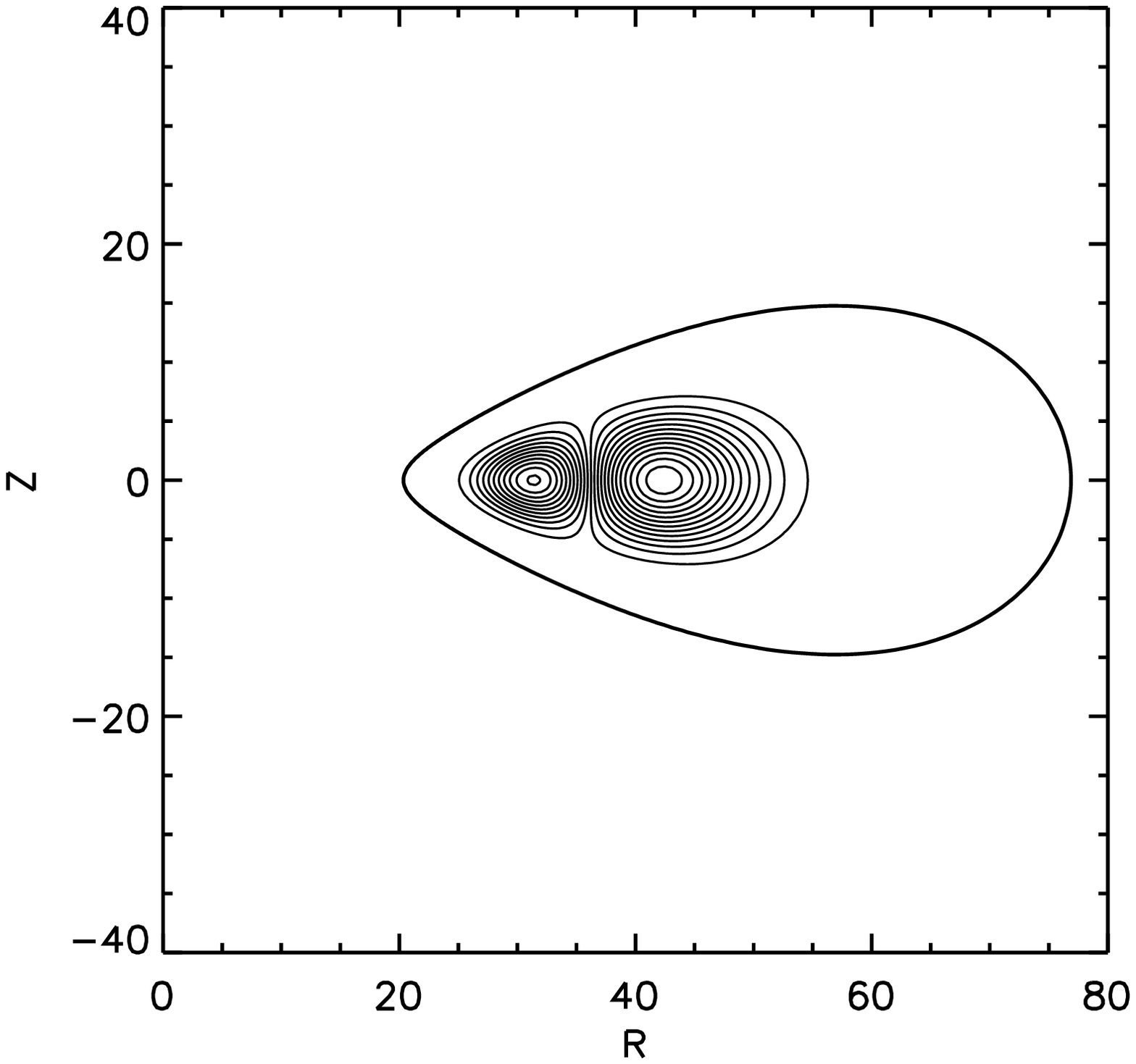}
\end{center}
\caption[]{Initial torus and field configuration for the two-loop
simulations.  The outermost contour line marks the boundary of the
initial torus.  The remaining contours are the magnetic field lines.}
\label{fig:initialfield}
\end{figure}

\begin{figure}
\leavevmode
\begin{center}
\includegraphics[width=0.7\textwidth]{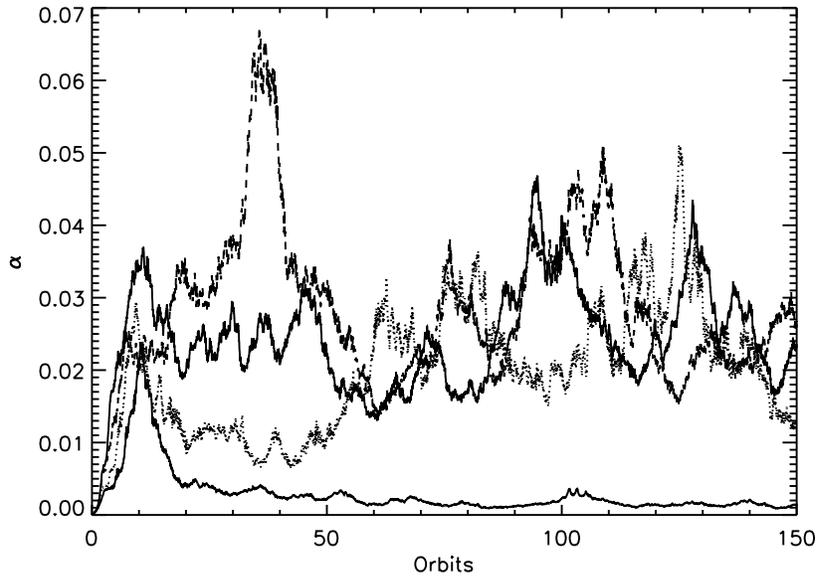}
\end{center}
\caption[]{Ratio of $\alpha$, the volume-averaged Maxwell and Reynolds stress to
volume-averaged pressure in a set of four stratified shearing box simulations.
The simulations use 8 (lowest solid line),
16 (dotted line), 32 (dashed line), and 64 zones (solid line) per 
scale-height $H$.  
}
\label{fig:shrstress}
\end{figure}

\begin{figure}
\leavevmode
\begin{center}
\includegraphics[width=0.7\textwidth,angle=90]{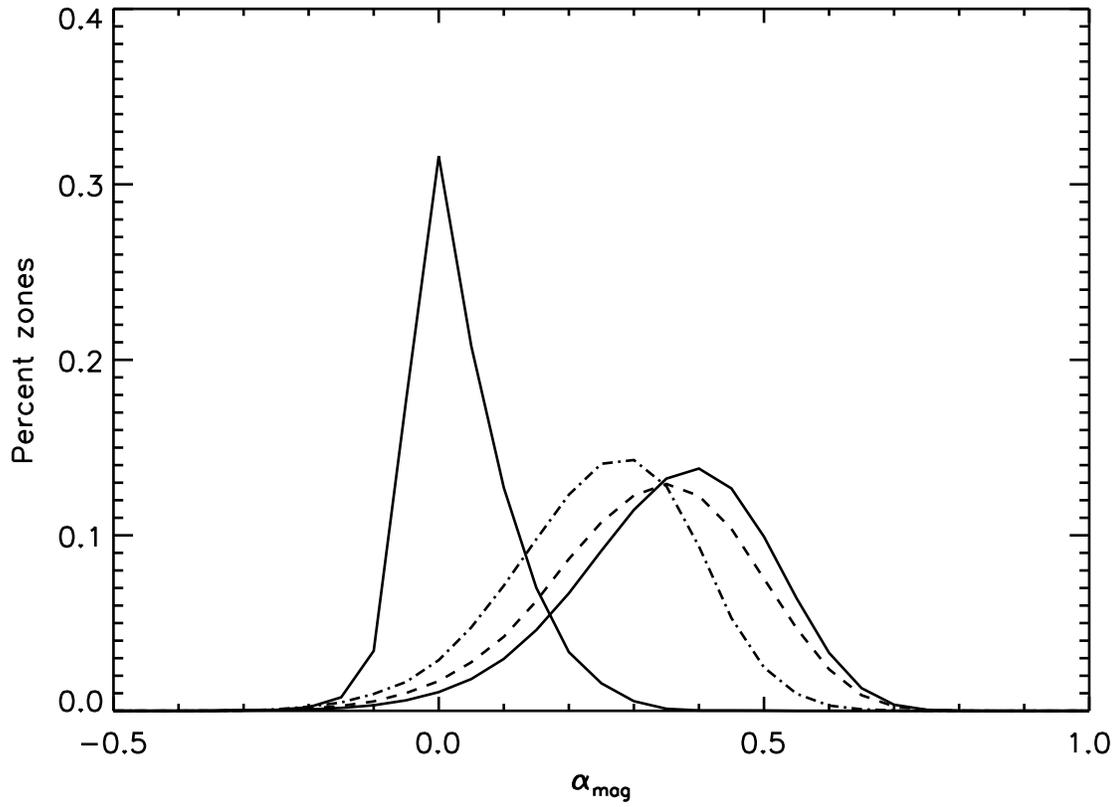}
\end{center}
\caption[]{Distribution function for the time- and volume-averaged
ratio of the Maxwell stress to the magnetic pressure, $\alpha_{mag}$
taken from a set of four stratified shearing box simulations using
8 (left-most solid line),
16 (dot-dashed line), 32 (dashed line), and 64 zones (solid line) per 
scale-height $H$.  As resolution increases the peak of the  distribution of
$\alpha_{mag}$ shifts to higher values.
}
\label{fig:alphamag}
\end{figure}

\begin{figure}
\leavevmode
\begin{center}
\includegraphics[width=0.7\textwidth]{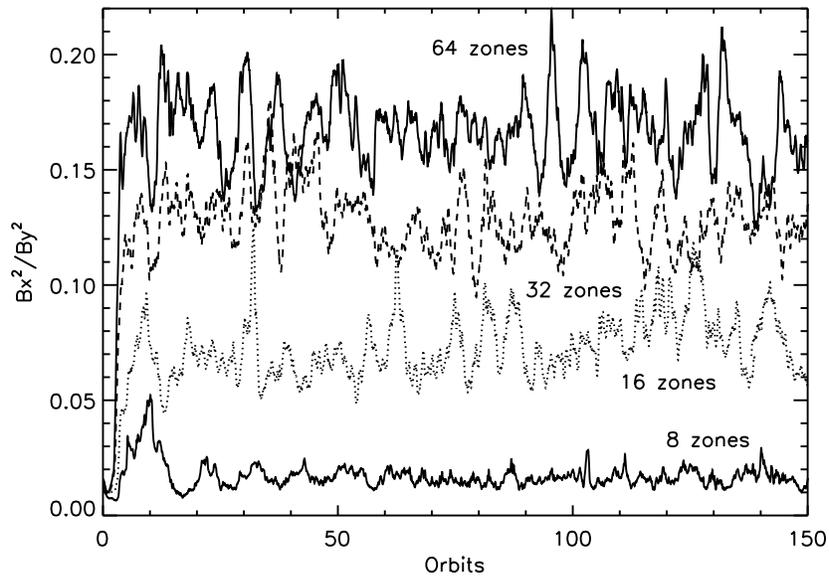}
\end{center}
\caption[]{Ratio of volume-averaged radial to toroidal field energies,
$\langle B_x^2\rangle/\langle B_y^2\rangle$ for stratified shearing
box simulations using 8 (solid line),
16 (dotted line), 32 (dashed line), and 64 zones (solid line) 
per scale-height $H$.  This ratio shows a
systematic increase with resolution.
}
\label{fig:shrbxby}
\end{figure}

\begin{figure}
\leavevmode
\begin{center}
\includegraphics[angle=90,width=0.7\textwidth]{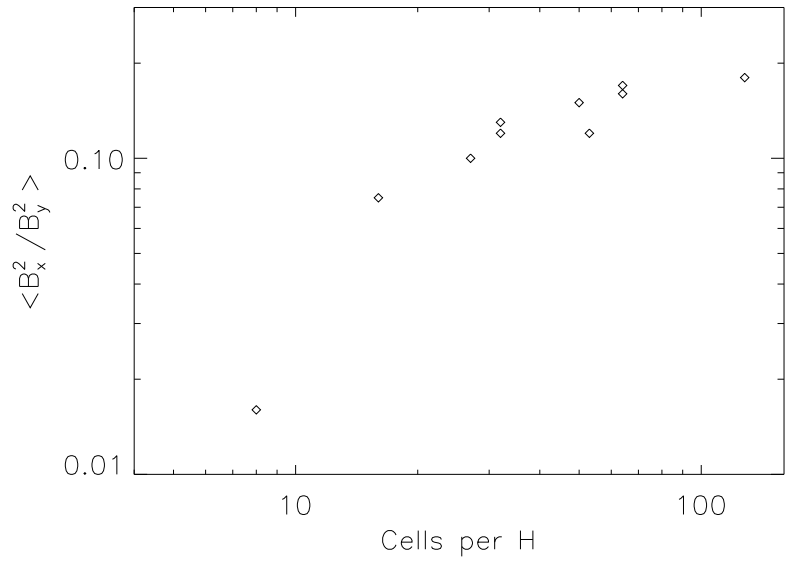}
\end{center}
\caption[]{Ratio of the volume-averaged radial field energy to
toroidal field energy, $\langle B_x^2/B_y^2\rangle$ as a function of resolution 
in stratified shearing box simulations.
Resolution is measured in number of grid cells per scale-height $H$.  
}
\label{fig:bxsqbysq}
\end{figure}

\begin{figure}
\leavevmode
\begin{center}
\includegraphics[width=0.7\textwidth]{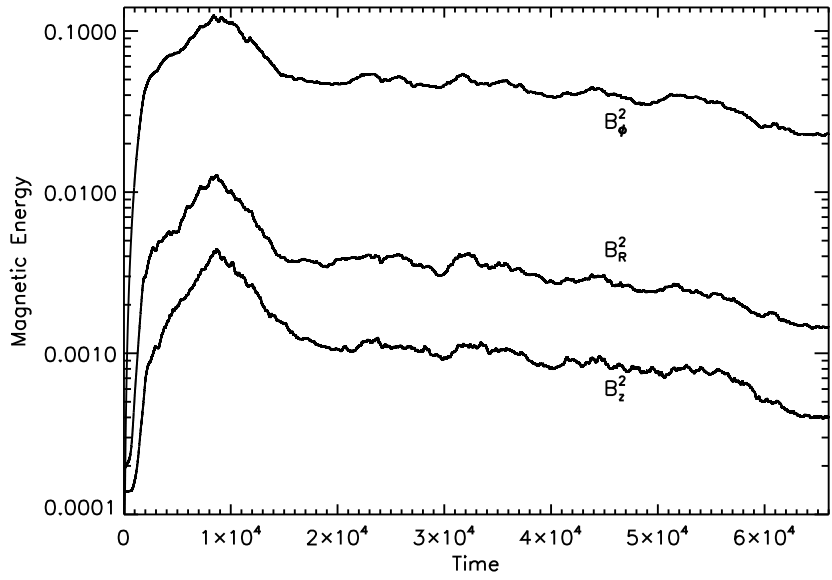}
\end{center}
\caption[]{Evolution of the three components of the magnetic energy as
a function of time in the fiducial simulation, twoloop-1000-mr.}
\label{fiducialenergy}
\end{figure}

\begin{figure}
\leavevmode
\begin{center}
\includegraphics[angle=90,scale=0.5]{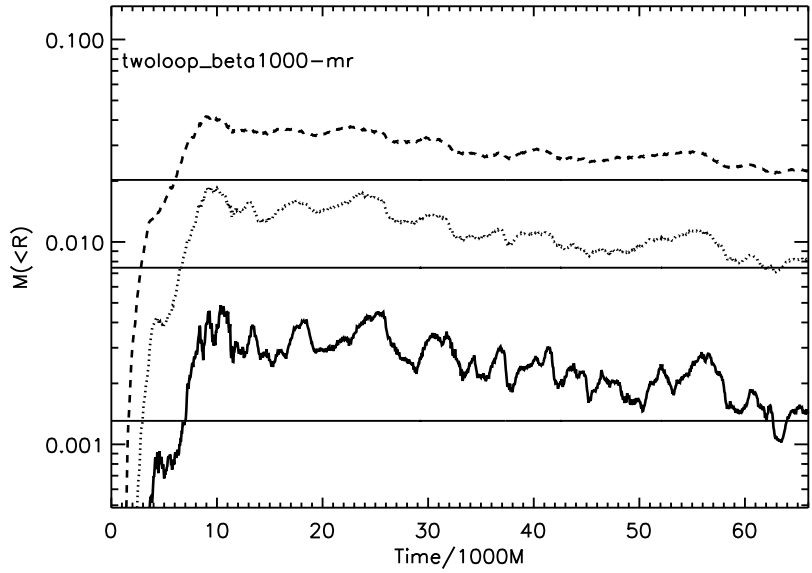}
\end{center}
\caption[]{The mass interior to three radii ($R=10M$: solid, $R=15M$: dotted,
$R=20M$: dashed) as a function of time in the fiducial simulation.
Mass units are fraction of the total initial mass.  The time averaged
accretion rate, $\dot M$, is 1\% of the initial mass every $3700 M$ in
time.  
}
\label{fig:massfillin}
\end{figure}

\begin{figure}
\leavevmode
\begin{center}
\includegraphics[angle=90,width=0.7\textwidth]{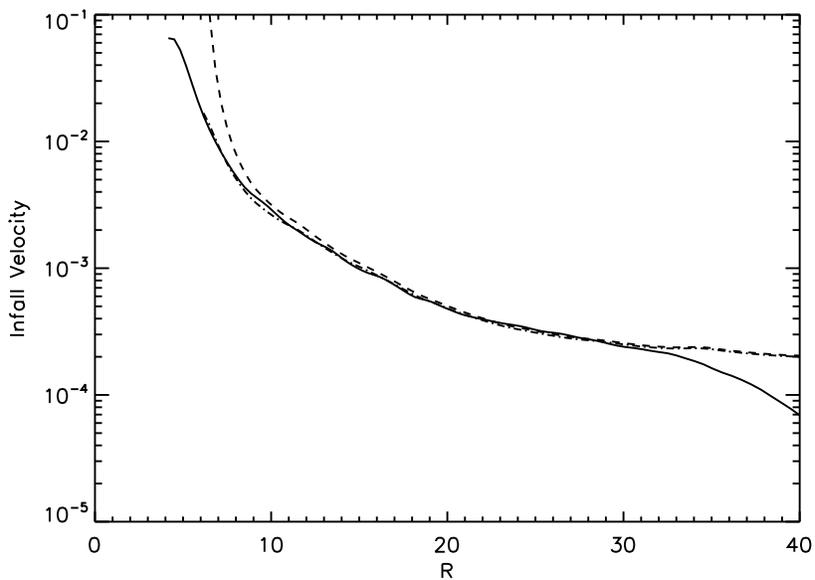}
\end{center}
\caption[]{
The inflow velocity for the fiducial run derived from the vertically and
azimuthally integrated accretion rate and density (solid line), along
with the infall velocity derived from the vertically and azimuthally
averaged Maxwell stress, pressure and accretion rate using a steady
state disk approximation (dashed line) with $j_* = j_{ISCO}$.  
The dot-dashed line is the steady state inflow velocity derived using
$j_* = 0.985 j_{ISCO}$.
The data are time-averaged from $t=4$--$5\times 10^4 M$.  
}
\label{fig:alphacomp}
\end{figure}

\begin{figure}
\leavevmode
\begin{center}
\includegraphics[width=0.7\textwidth]{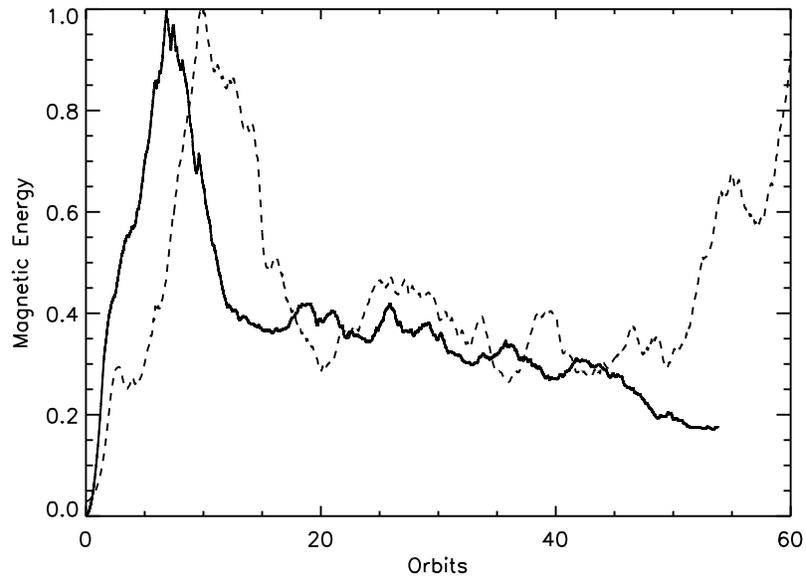}
\end{center}
\caption[]{Evolution of the magnetic energy in the fiducial run (solid line)
compared to the magnetic energy in a stratified shearing box
simulation with 16 zones per $H$ (dashed line).
Time is orbits at the initial torus pressure maximum or in terms of
the shearing box $\Omega$,  and the magnetic energies are
normalized to their peak value.  Both models show a
period of rapid field amplification followed by a peak and a decline
to a longer-term value that slowly declines.  The magnetic energy increases 
after 50 orbits in the shearing box, but not in the global model.
}
\label{fig:stratcomp}
\end{figure}

\begin{figure}
\leavevmode
\begin{center}
\includegraphics[angle=90,scale=0.4]{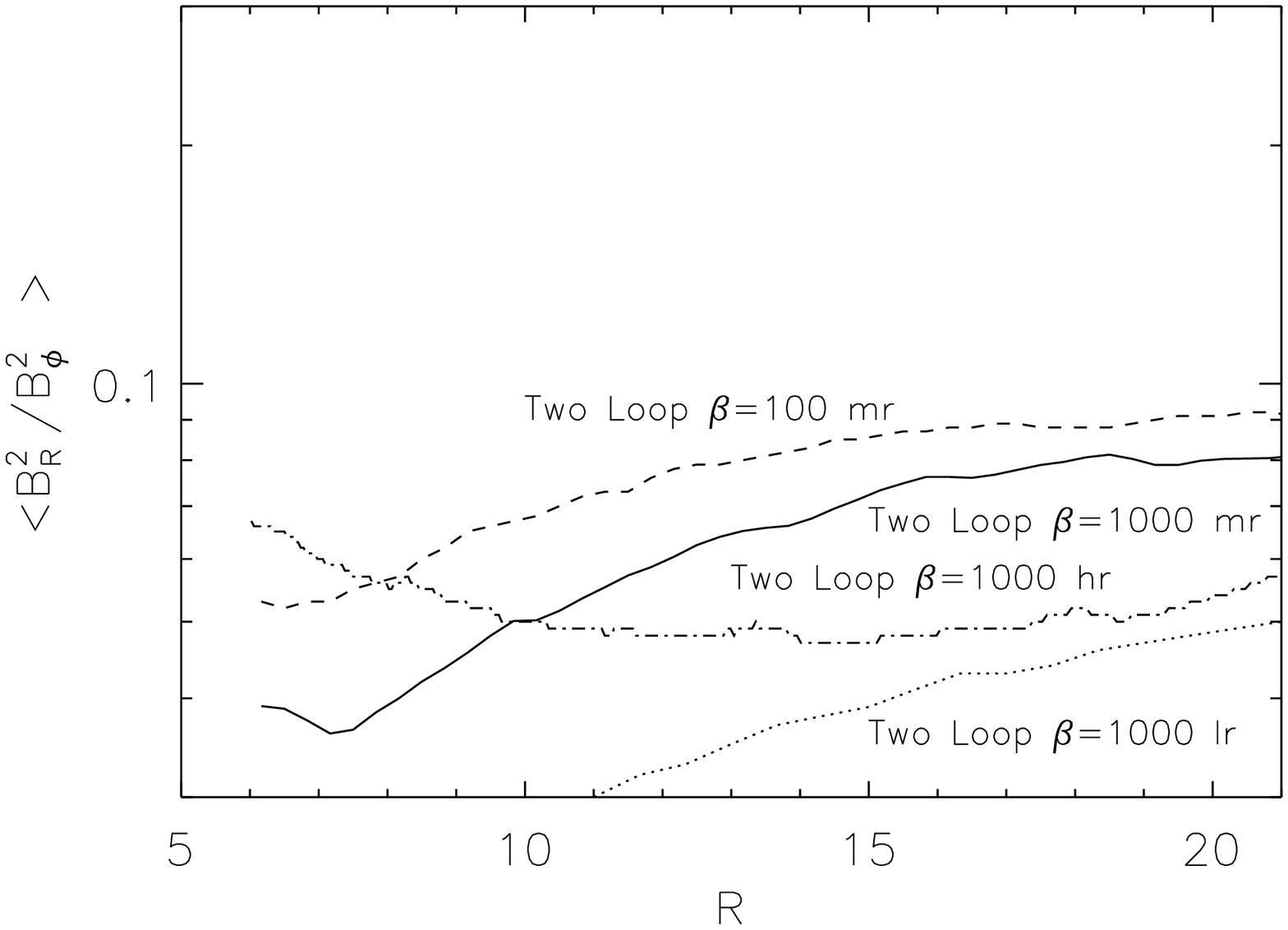}
\includegraphics[angle=90,scale=0.4]{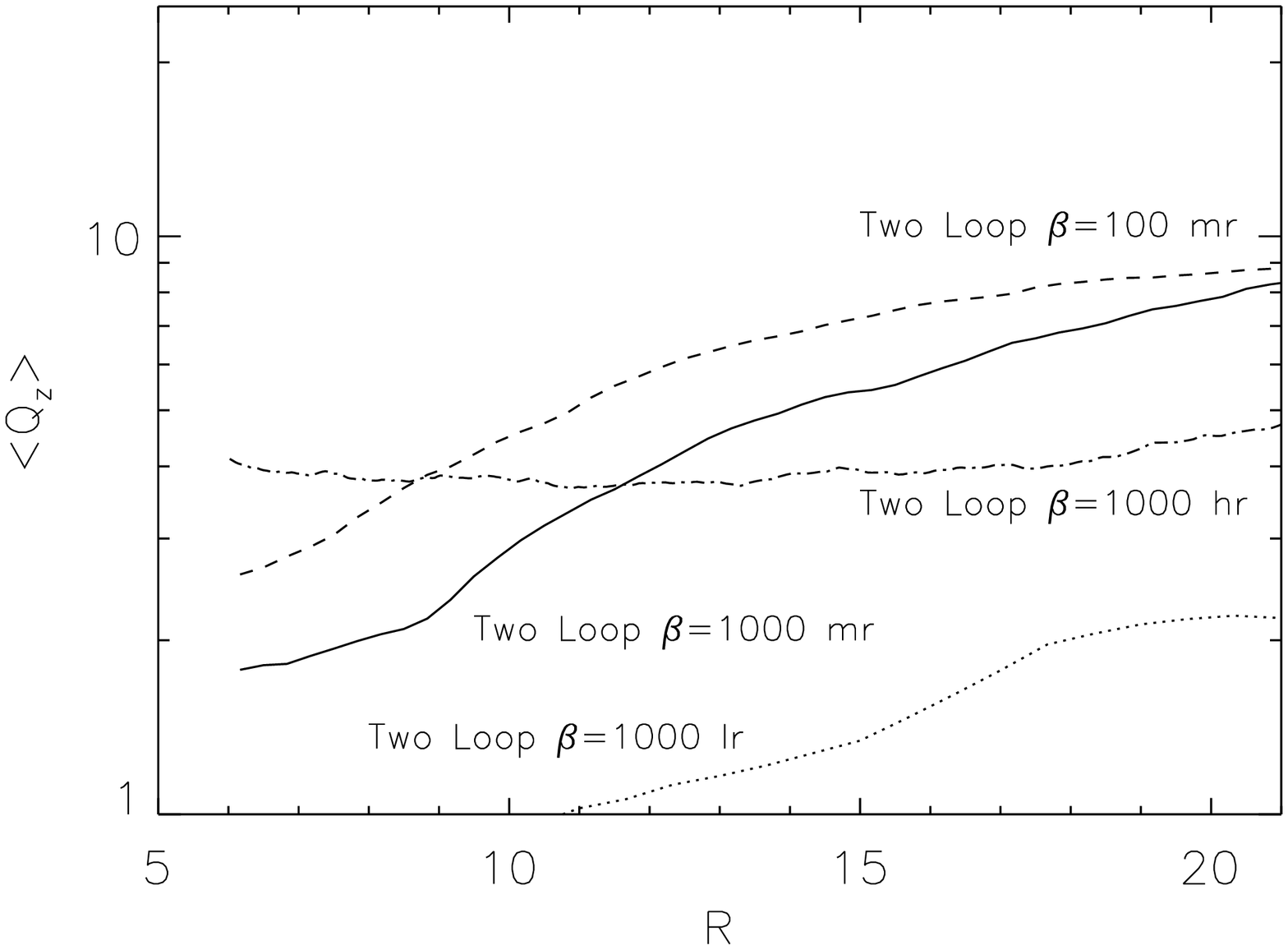}
\includegraphics[angle=90,scale=0.4]{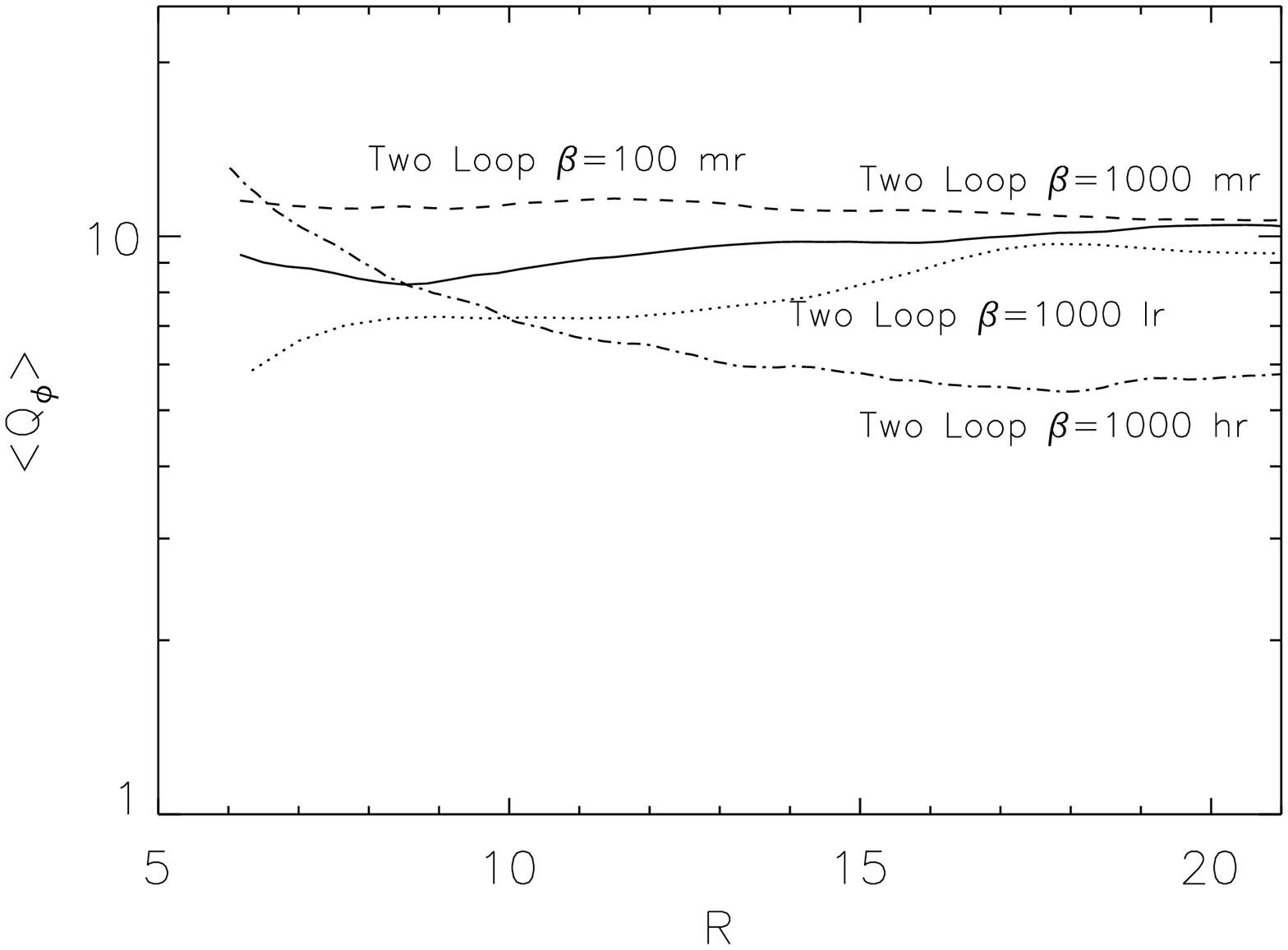}
\end{center}
\caption[]{
Radial dependence of the principal diagnostics for the 2-loop simulations
with the standard azimuthal resolution.  The curves are labeled with
model name.  The data were averaged over
azimuth and height with a density weighting and then averaged in
time from $2$--$3.4\times 10^4 M$ for all
three simulations.  (Top) $\brsq$, (Middle) $\Qz$, (Bottom) $\Qp$.
}
\label{fig:twoloopdiags}
\end{figure}

\begin{figure}
\leavevmode
\begin{center}
\includegraphics[angle=90,scale=0.4]{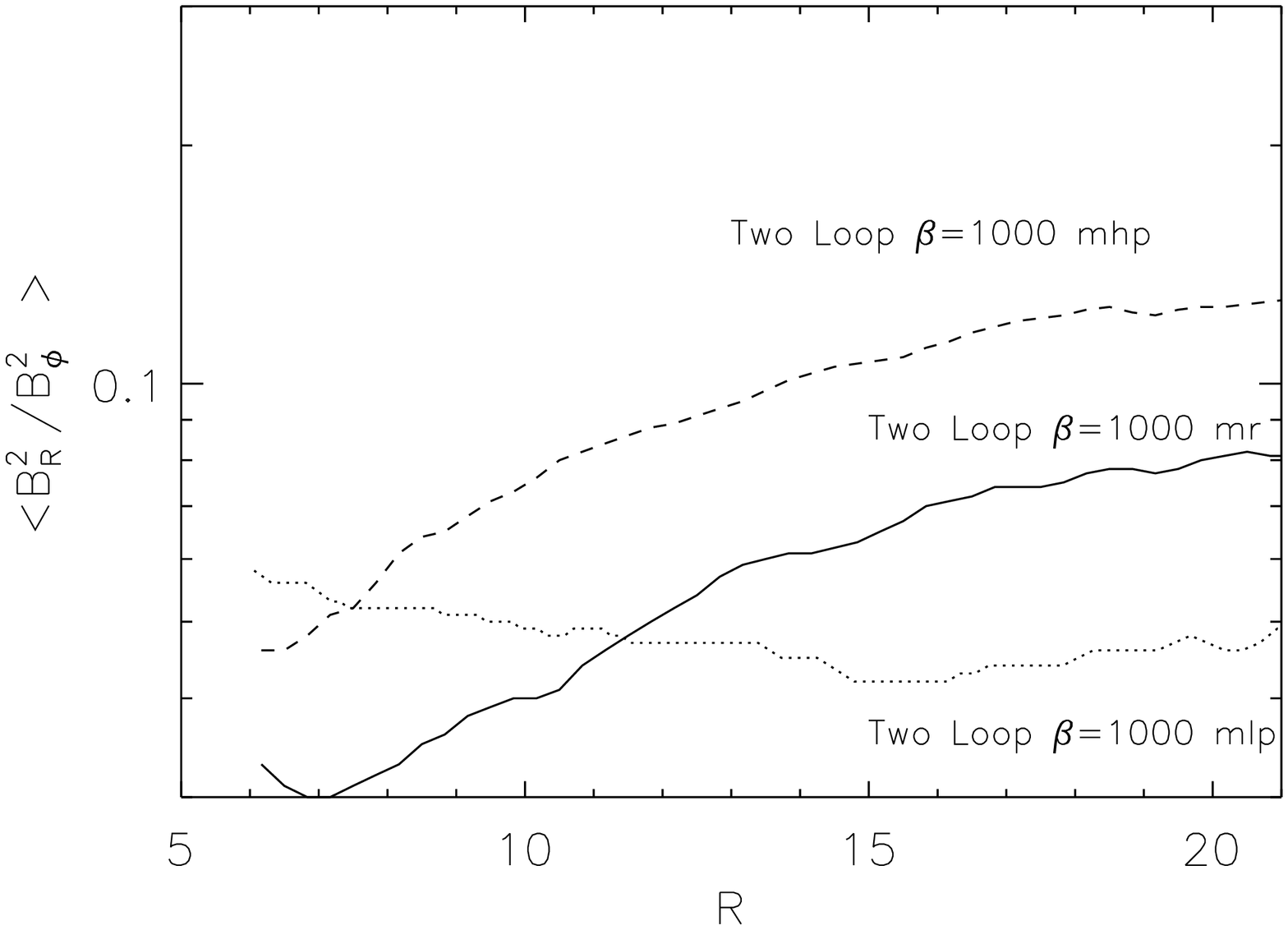}
\includegraphics[angle=90,scale=0.4]{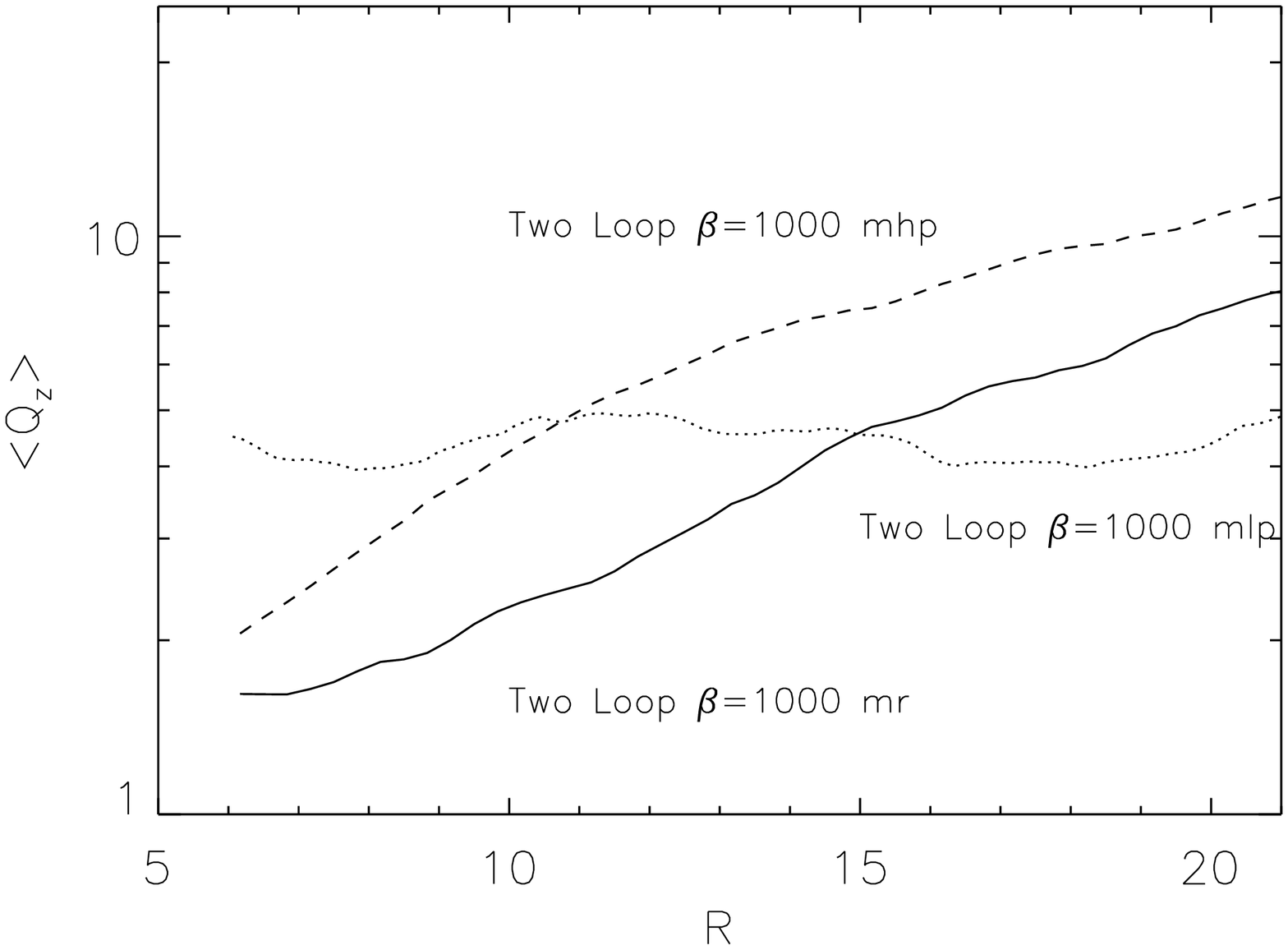}
\includegraphics[angle=90,scale=0.4]{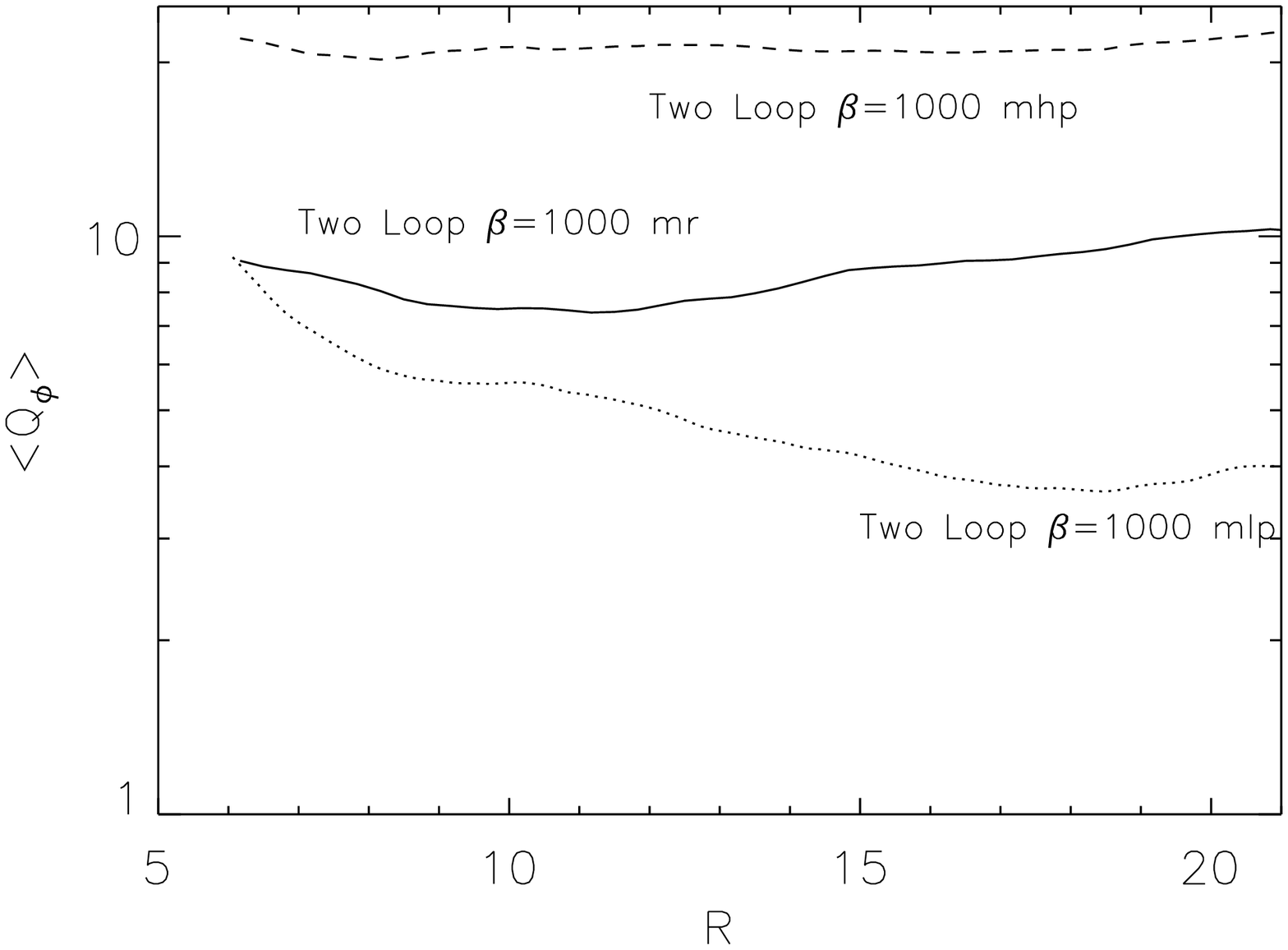}
\end{center}
\caption[]{
Radial dependence of the principal diagnostics for the 2-loop simulations
with varying azimuthal resolution.  The lines are labeled with model
name.  The data were averaged over azimuth and
height with a density weighting and then averaged in time from
$1.8$--$2.5\times 10^4 M$ for all 
three simulations.  (Top) $\brsq$, (Middle) $\Qz$, (Bottom) $\Qp$.
}
\label{fig:phiresdiags}
\end{figure}

\begin{figure}
\leavevmode
\begin{center}
\includegraphics[width=0.7\textwidth]{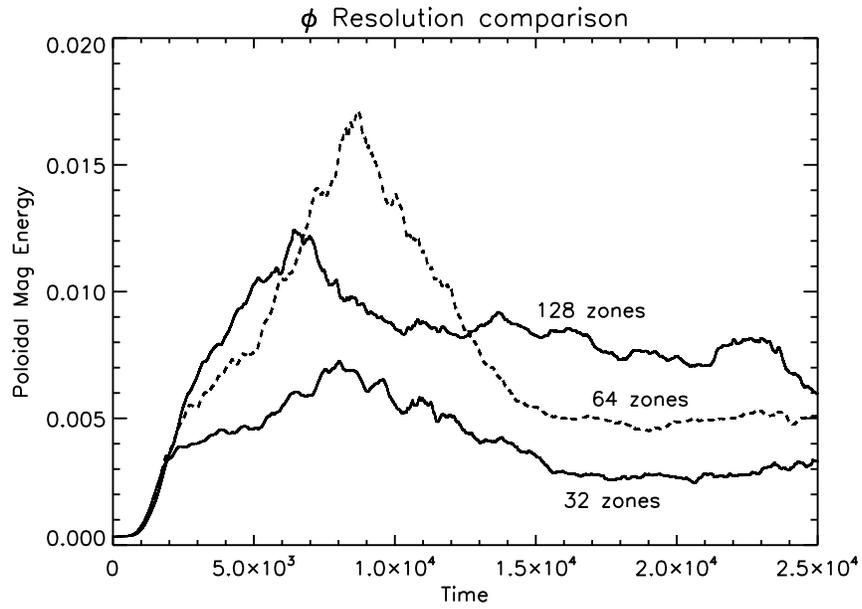}
\end{center}
\caption[]{
Evolution of three simulations using different numbers of zones to
span the $\phi$ domain:  32, 64 (dashed line) and 128.  The initial
conditions were the same: two magnetic field loops with average strength
$\beta = 1000$.
}
\label{fig:comphi}
\end{figure}

\begin{figure}
\leavevmode
\begin{center}
\includegraphics[angle=90,scale=0.4]{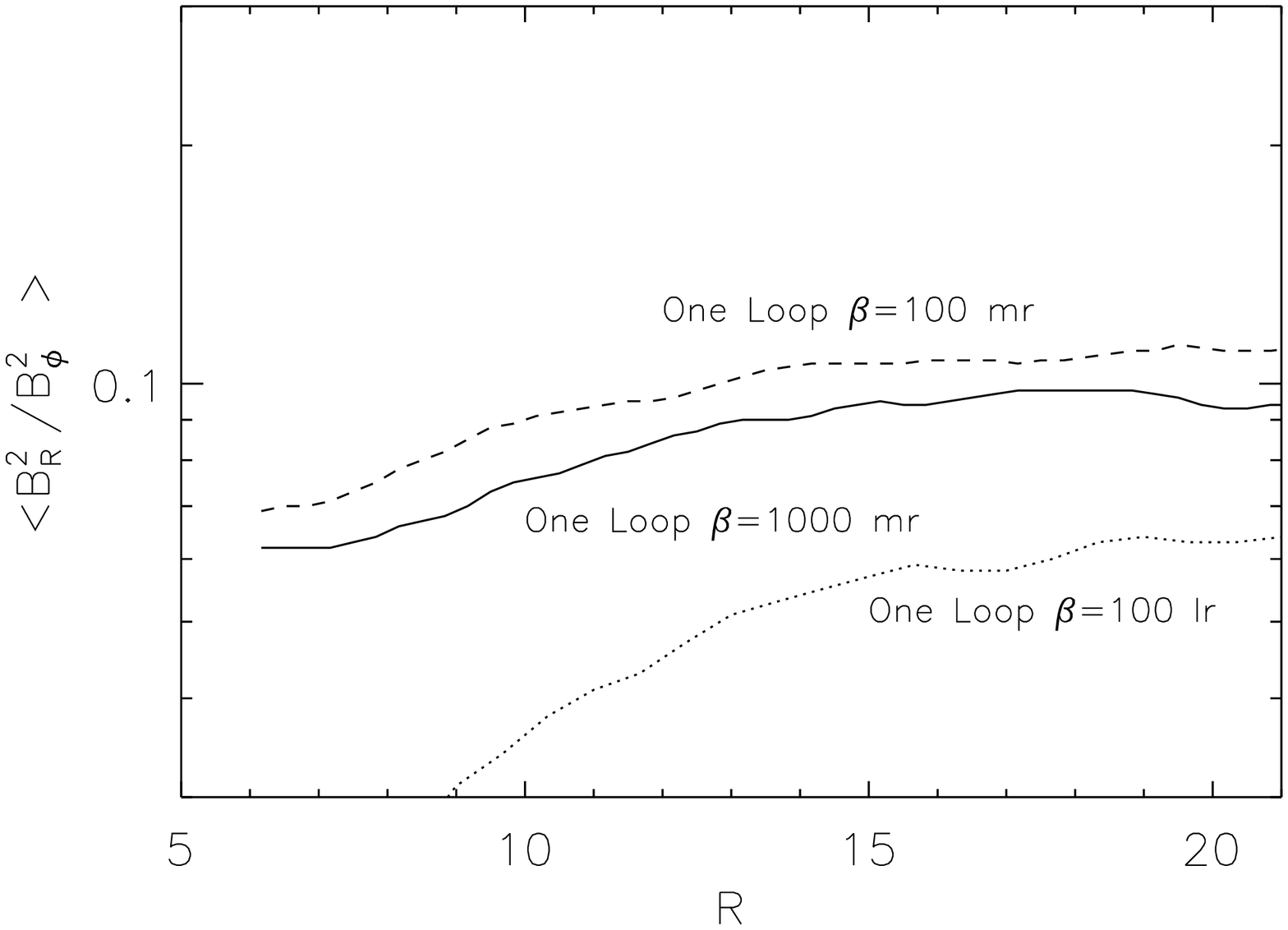}
\includegraphics[angle=90,scale=0.4]{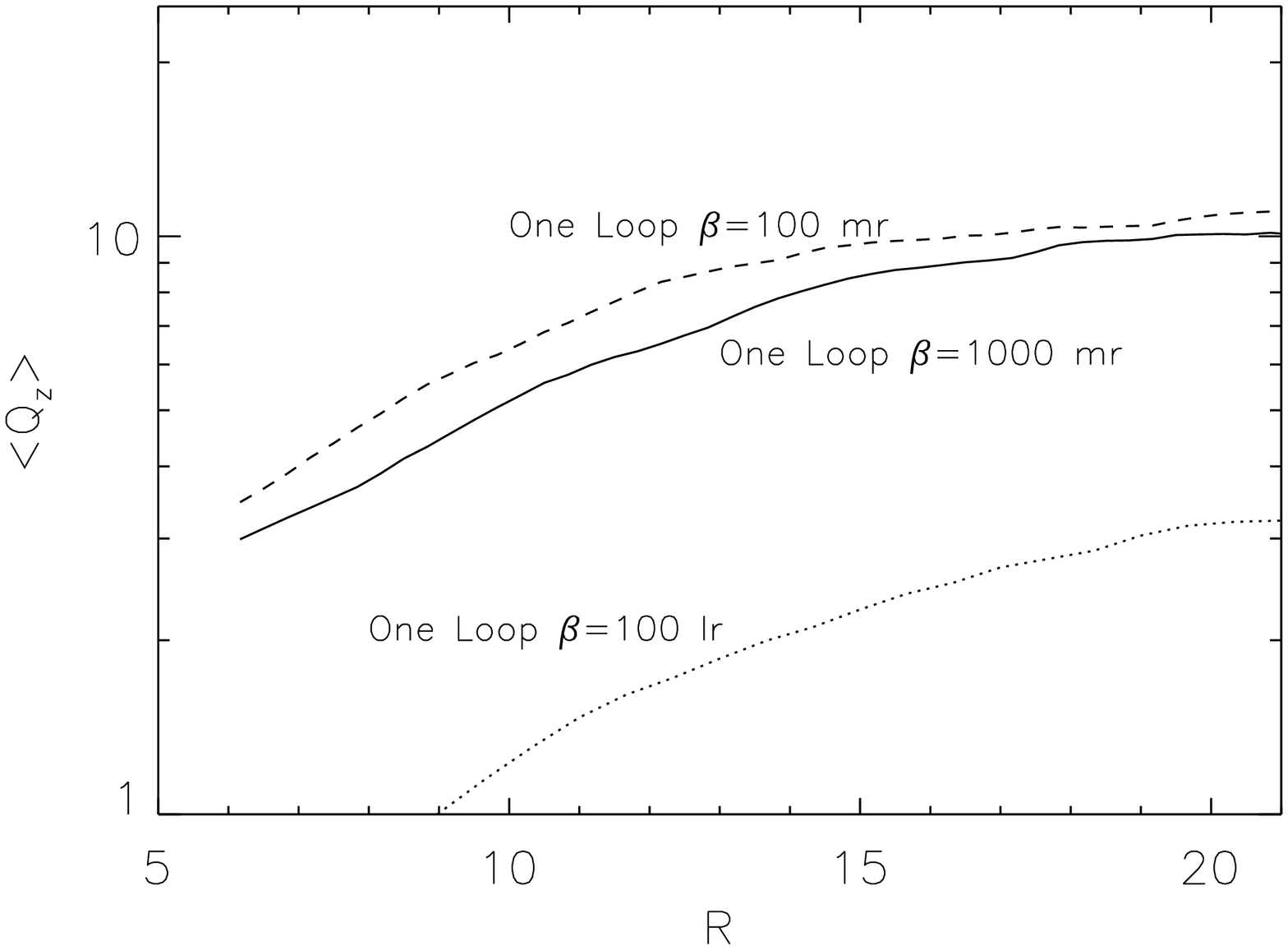}
\includegraphics[angle=90,scale=0.4]{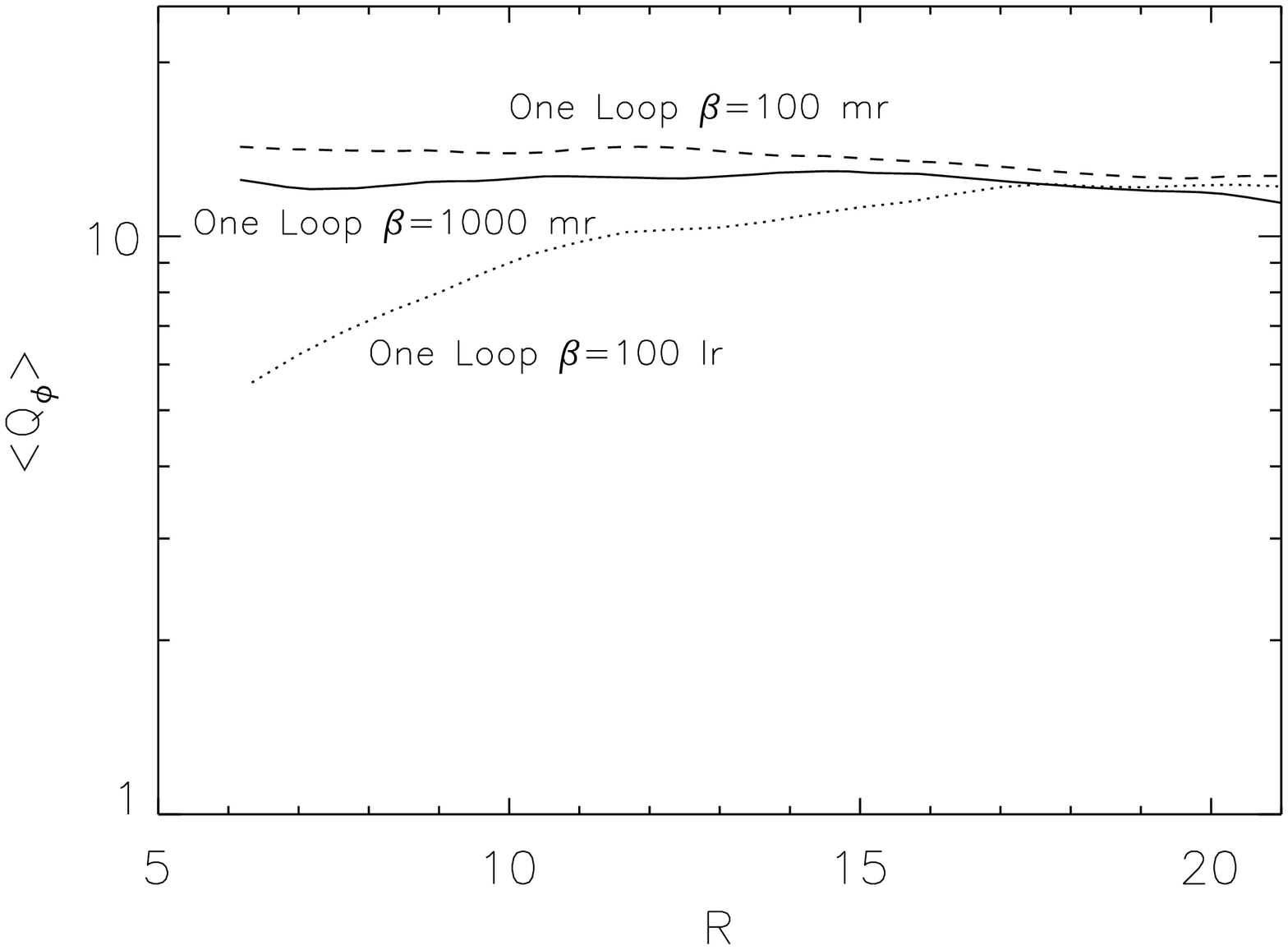}
\end{center}
\caption[]{
Radial dependence of the principal diagnostics for the 1-loop simulations
with the standard azimuthal resolution.  The curves are labeled with
the corresponding model name.  The data were averaged over
azimuth and height with a density weighting and then averaged in
time from $2$--$3.8 \times 10^4 M$ for all
three simulations.  (Top) $\brsq$. (Middle) $\Qz$.  (Bottom) $\Qp$.
}
\label{fig:oneloopdiags}
\end{figure}

\begin{figure}
\leavevmode
\begin{center}
\includegraphics[angle=90,width=0.7\textwidth]{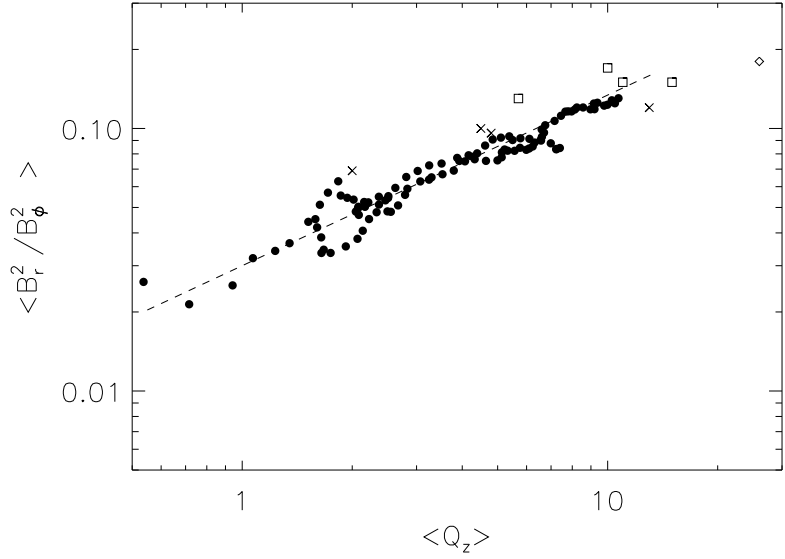}
\end{center}
\caption[]{
Correlation between $\Qz$ and $\brsq$. The points shown by
filled circles in this figure are drawn from the same set of times in
three different simulations: twoloop-beta1000-lr, twoloop-beta1000-mr
(the fiducial global simulation), and twoloop-beta1000-mhp.  All are
density-weighted averages, but they are taken at different radii within
the inner disk.  The points shown by the other symbols are time-averages
of data from the midplane region of the shearing box simulations.
The different symbols within this category denote different ranges of
$\Qy$: $12\le \Qy \le 25$ (x's); $25 < \Qy \le 50$ (squares); $\Qy=98$
(diamond).  The dashed line has a slope of 0.65.
}
\label{fig:qzvsbrby}
\end{figure}

\begin{figure}
\leavevmode
\begin{center}
\includegraphics[angle=90,width=0.7\textwidth]{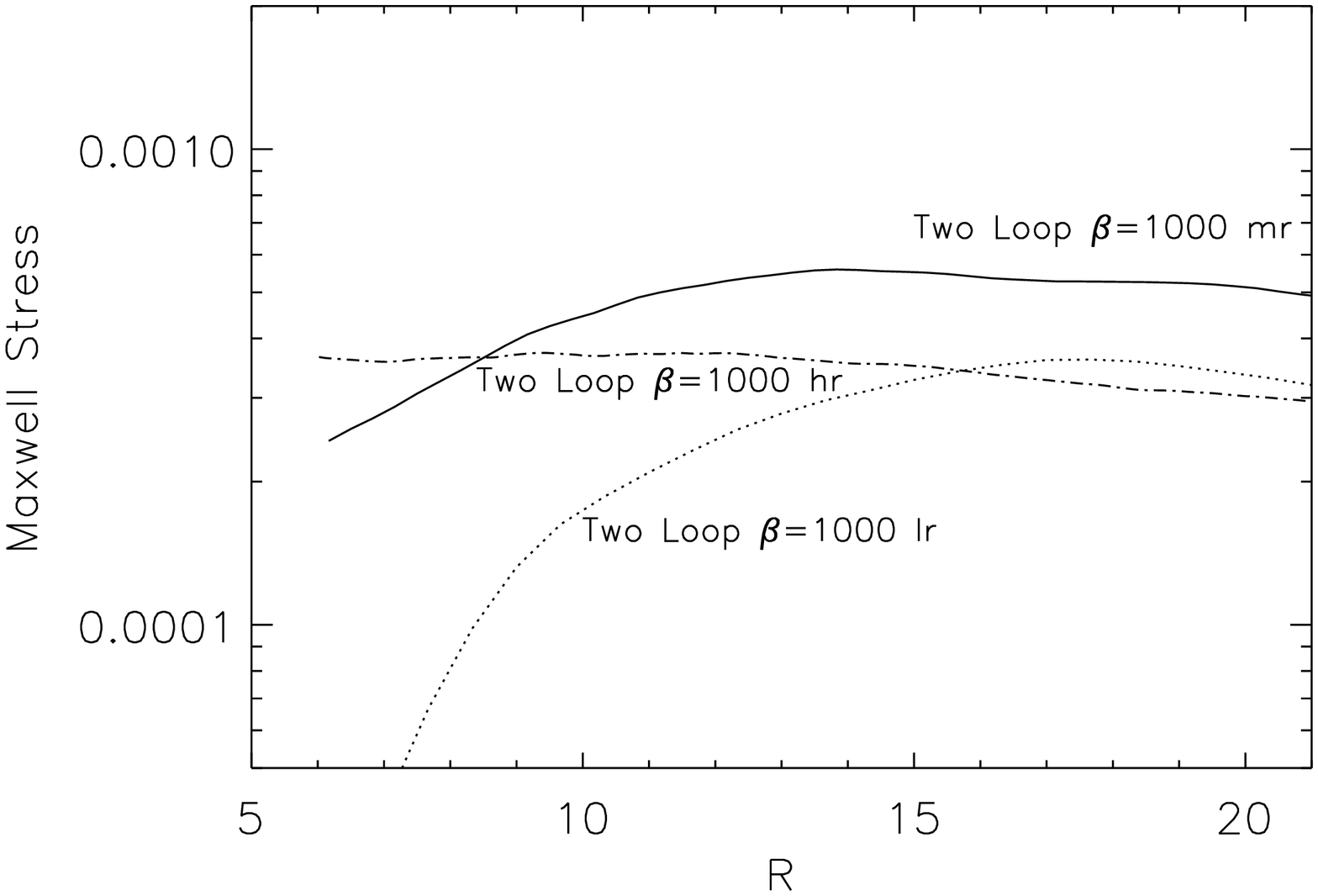}
\includegraphics[angle=90,width=0.7\textwidth]{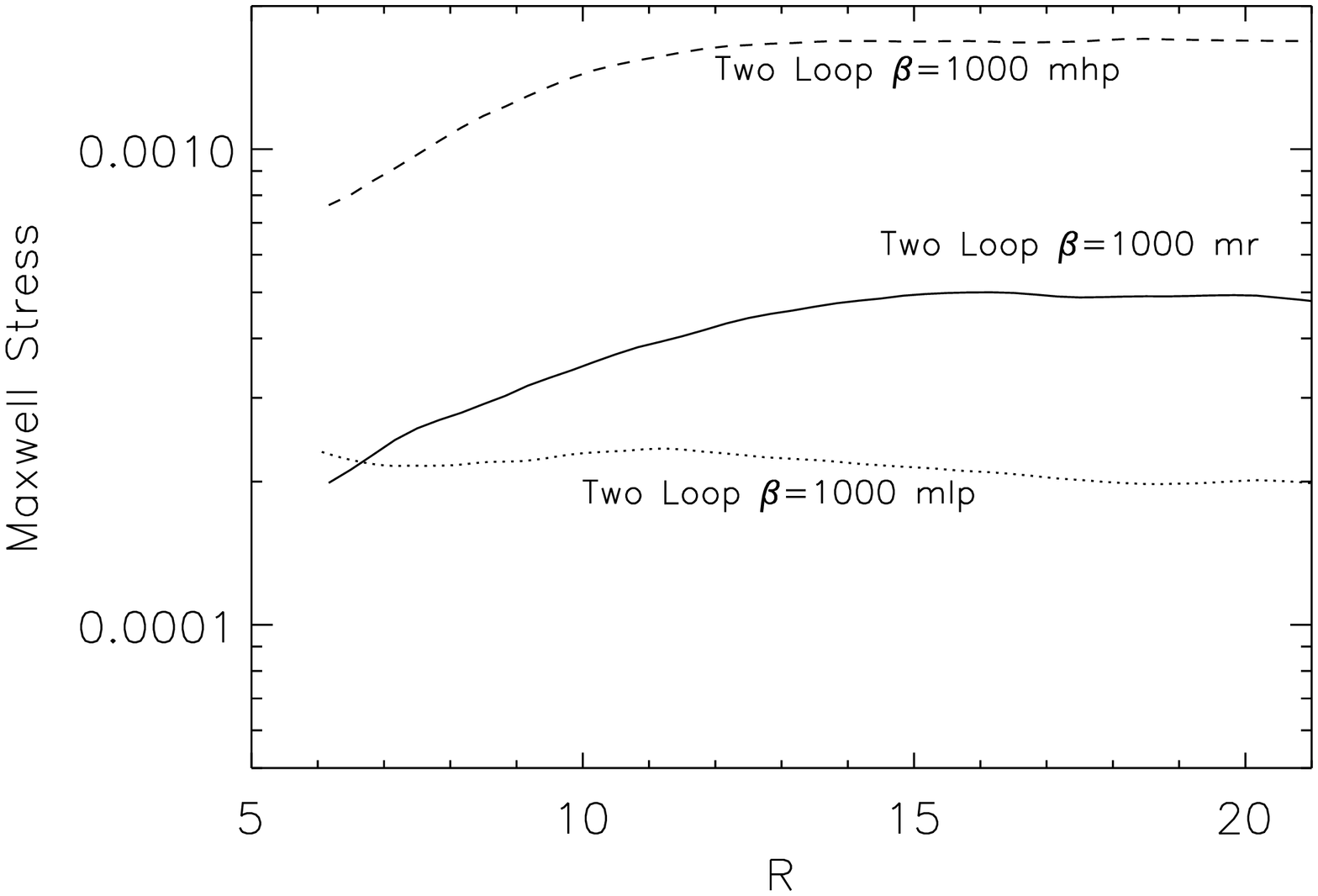}
\end{center}
\caption[]{
Time-averaged and shell-integrated Maxwell stress profiles for the three
2-loop simulations differing only in poloidal grid (Top) and the three
2-loop simulations differing only in azimuthal grid (Bottom).  Curves
are labeled by their model name.
}
\label{fig:stressprofcomp}
\end{figure}